\documentclass{article} 
\usepackage{amssymb}

\def\baselinestretch{1.5}
\begin{document}
\newcommand{\gen}[1]{\mbox{$\langle #1 \rangle$}}
\newcommand{\mchoose}[2]{\def\baselinestretch{1} \left ( {
 {#1} \choose{#2} } \right ) \def\baselinestretch{2}}
\def\Box{[]}
\newenvironment{proof}{\noindent{\sc Proof:\ }}{\vspace{2ex}}
\newenvironment{prooft}{{\bf

Proof of the Theorem.}}{$\Box$ \vspace{1ex}}
\newenvironment{prooff}{{\bf

Proof.}}{\vspace{1ex}}
\newenvironment{proofl}{{\bf
Proof.}}{$\Box$ \vspace{1ex}}


\def\numberlikeadb{\global\def\theequation{\thesection.\arabic{equation}}}
\numberlikeadb
\newtheorem{theorem}{Theorem}[section]
\newtheorem{lemma}[theorem]{Lemma}
\newtheorem{corollary}[theorem]{Corollary}
\newtheorem{proposition}[theorem]{Proposition}
\newtheorem{example}[theorem]{Example}

\newcommand{\hh}{{\hspace{.3cm}}}
\newcommand{\RR}{{\bf R}}
\newcommand{\NN}{{\bf N}}
\newcommand{\ZZ}{{\bf Z}}
\newcommand{\PP}{{\bf P}}
\newcommand{\EE}{{\bf E}}
\newcommand{\var}{{\mbox{Var}}}
\newcommand{\Cov}{{\mbox{Cov}}}
\newcommand{\beas}{\begin{eqnarray*}}
\newcommand{\enas}{\end{eqnarray*}}
\newcommand{\fone}{f_1^{(1)}}
\newcommand{\ftwo}{f_1^{(2)}}
\newcommand{\wnone}{W(n)^{(1)}}
\newcommand{\wntwo}{W(n)^{(2)}}

\newcommand{\ctilde}{{\tilde {c}}}

\newcommand{\bea}{\begin{eqnarray}}
\newcommand{\ena}{\end{eqnarray}}
\newcommand{\eq}{\begin{equation}}
\newcommand{\en}{\end{equation}}
\newcommand{\G}{\frac{\g}{4-\g}}

\def\ignore#1{}
\def\comm#1{\footnote{#1 }}

\def\gcomm#1{\comm{(Gesine)---#1}}
\def\grcomm#1{\comm{(Gesine)---Could not read this word#1}}
\def\acomm#1{\comm{(Andrew)---#1}}

\def\half{{\textstyle{\frac12}}}
\def\be{{\bf e}}
\def\uo{^{(0)}}
\def\ul{^{(l)}}
\def\bM{{\bf M}}
\def\bW{{\bf W}}
\def\Ref#1{(\ref{#1})}
\def\a{\alpha}
\def\b{\beta}
\def\s{\sigma}
\def\f{\phi}
\def\w{\omega}
\def\l{\lambda}
\def\L{\Lambda}
\def\tod{\buildrel{{\cal D}{\to}}}
\def\eqd{\buildrel{{\cal D}{=}}}
\def\pnmw{\PP_{NMW}}
\def\whv{{\widehat V}}
\def\fij{\phi_{ij}}
\def\Xij{X_{ij}}
\def\pij{p_{ij}}
\def\siim{\sum_{i=1}^m}
\def\sjn{\sum_{j=1}^n}
\def\dtv{d_{TV}}
\def\law{{\cal L}}
\def\Zij{Z_{ij}}
\def\ep{\hfil $\Box$ 
\def\hm{{\hat M}} 
\def\hn{{\hat N}} 

\bigskip}
\def\Tr#1{Theorem~\ref{#1}}
\def\giv{\,|\,}
\def\remark{\noindent {\bf Remark.}\ \,}
\def\Yil{Y_{il}}
\def\tfrac#1#2{{\textstyle{\frac#1#2}}}

\def\Po{{\rm Po\,}}
\def\BHJ{\cite{bhj}}
\def\non{\nonumber}
\def\th{\theta}
\def\bN{{\bf N}}
\def\e{\varepsilon}
\def\m{\mu}
\def\uil{^{(il)}}
\def\var{{\rm Var\,}}
\def\dd{\delta}
\def\bn{{\bigskip \noindent}}
\def\r{\rho}
\def\etal{{\it et.~al.\/}}
\def\t{\tau}
\def\g{\gamma}
\def\h{\eta}
\def\f{\phi}
\def\siiM{\sum_{i=1}^M}
\def\sjN{\sum_{j=1}^N}
\def\ps{\psi}
\def\lnti{{\lim_{n\to\infty}}}
\def\un{^{(n)}}
\def\Bl{\left(}
\def\Br{\right)}
\def\Blm{\left|}
\def\Brm{\right|}
\def\Blb{\left\{}
\def\Brb{\right\}}
\def\NE{{\rm NE\,}}
\def\ui{^{(1)}}
\def\ut{^{(2)}}
\def\tu{{\widetilde U}}
\def\hs{{\hat s}}
\def\Bi{{\rm Bi\,}}
\def\k{\kappa}
\def\uid{'{}\ui}
\def\hu{{\hat u}}
\def\nin{\noindent}
\def\D{\Delta}
\def\oper{\Psi}
\def\messls{{\Bl\frac\l{\l-\l_2}\Br^2(\l-\s)(2\l-\s)}}
\def\fiimi{(f_1\ui)^{-1}}
\def\fii{{f_1\ui}}
\def\lmit{{\tfrac{{\l-1}}{2}}}
\def\ex{{\bf E}}
\def\pr{{\phi_\r}}

\title{Discrete small world networks}
\author{
A. D. Barbour\footnote{
ADB was supported in part by Schweizerischer Nationalfonds Projekt Nr.\
20-67909.02.}\\
Applied Mathematics\\
University of Z\"urich\\
CH - 8057 Z\"urich\\
and
\\
G. Reinert\footnote{
GR was supported in part by EPSRC grant no. GR/R52183/01.}\\
Department of Statistics\\
University of Oxford\\
UK - Oxford OX1 3TG}

\date{} 
\maketitle

\begin{abstract}
Small world models are networks consisting of many local links and 
fewer long range `shortcuts', used to model networks with a high 
degree of local clustering but relatively small diameter. Here, 
we concern ourselves with the distribution of typical 
inter-point network distances. We establish approximations to the
distribution of the graph distance in a discrete ring network
with extra random links, and compare the results to those for
simpler models, in which the extra links have zero length and 
the ring is continuous. 
\end{abstract}

\section{Introduction}
 \setcounter{equation}{0}

There are many variants of the mathematical model  introduced  
by Watts and Strogatz~\cite{WattsStrogatz}
to describe the ``small--world'' networks
popular in the social sciences; one of them, the great circle
model of Ball~\etal~\cite{Balletal}, actually 
precedes~\cite{WattsStrogatz}.  See \cite{AlbertBarabasi} for a 
recent overview, as well as the books \cite{barabasi} and \cite{dormenbook}.  
A typical description is as follows.
Starting from a ring lattice with~$L$ vertices, 
 each vertex is  connected to all of its neighbours within
distance~$k$ by an undirected edge.  Then a number of shortcuts are added
between randomly chosen pairs of sites. Interest centres on the
statistics of the shortest distance between two (randomly chosen) vertices,
when shortcuts are taken to have length zero. 

Newman, Moore  and Watts \cite{NewmanWatts},
\cite{NewmanWatts2} proposed an idealized version, in which
the lattice is replaced by a circle and distance along the circle
is the usual arc length, shortcuts now being added between 
random pairs of uniformly distributed points.  Within their~[NMW] model,
they made a heuristic computation of the mean distance between a
randomly chosen pair of points.  Then Barbour and Reinert~\cite{BR}
proved an asymptotic approximation for the distribution of this
distance as the mean number $L\r$ of shortcuts tends to infinity;
the parameter~$\r$ describes the average intensity of end points
of shortcuts around the circle.
In this paper, we move from
the continuous model back to a genuinely discrete model, in which
the ring lattice consists of exactly~$L$ vertices, each with
connections to the~$k$ nearest neighbours on either side, but in
which the random shortcuts, being edges of the graph, are taken
to have length~$1$; thus distance becomes the usual graph distance
between vertices.  However, this model is rather complicated to
analyze, so we first present a simpler version, in which time
runs in discrete steps, but the process still lives on the
continuous circle, and which serves to illustrate the main
qualitative differences between discrete and continuous models.
This intermediate model would be reasonable for describing the
spread of a simple epidemic, when the incubation time of the
disease is a fixed value, and the infectious period is very
short in comparison.   In each of these more complicated models, 
we also show that the approximation derived for the [NMW] model
gives a reasonable approximation to the 
distribution of inter-point distances, 
provided that~$\r$ (or its equivalent) is small; here,  
the error in Kolmogorov distance is of order 
$O(\rho^\frac{1}{3} \log(\frac{1}{\rho}))$, although the distribution 
functions are only $O(\rho)$ apart in the bulk of the distribution.

\section{The continuous circle model for discrete time}\label{Sect2}  
\setcounter{equation}{0}  

In this section, we consider the continuous model of~\cite{BR}, 
which consists of a 
circle $C$ of circumference $L$, to which are added a Poisson 
$\Po(L\rho/2)$ number of 
uniform and independent random chords, but now with a new measure 
of distance between
points $P$ and~$Q$. This distance is the minimum of~$d(\g)$ 
over paths~$\g$ along the
graph between $P$ and~$Q$, where, if~$\g$ consists of~$s$
arcs of lengths $l_1,\ldots,l_s$ connected by shortcuts, then
$d(\g) := \sum_{r=1}^s \lceil l_r\rceil$, where, as usual,
$\lceil l\rceil$ denotes the smallest integer $m\ge l$;
shortcuts make no contribution to the distance.
We
are interested in asymptotics as $L\rho \to \infty$, and so
assume throughout that $L\rho > 1$.

We begin with a dynamic realization of 
 the network, which describes, for each $n\ge0$, the set of points $R(n)
\subset C$ that can be reached from a given point~$P$ within time~$n$,
where time corresponds to the~$d(\cdot)$ distance along paths.  
Pick Poisson $\Po(L\rho)$ uniformly and
independently distributed `potential' chords of the circle $C$;
such a chord is an unordered pair of independent and uniformly 
distributed random points of~$C$.  Label one point of each pair
with~$1$ and the other with~$2$, making the choices equiprobably,
independently of everything else.  We
call the set of label~1 points~$Q$, and, for each $q \in Q$,
we let $q'=q'(q)$ denote the label~2 end point. Our construction
realizes a random subset of these potential chords as shortcuts.
We start by taking $R(0)=\{P\}$ and $B(0)=1$,
and let time increase in integer steps.
$R(n)$ then consists of a union of~$B(n)$ intervals of~$C$, 
each of which is
increased by unit length at each end point at time~$n+1$, 
but with the rule that
overlapping intervals are merged into a single interval; 
this defines a new union of~$B'(n+1)$ intervals~$R'(n+1)$;
note that~$B'(n+1)$ may be less than~$B(n)$. 

Now define $\partial R(n+1) := R'(n+1) \setminus R(n)$. 
Whenever $\partial R(n+1) \cap Q$ is not empty --- 
that is, whenever $\partial R(n+1)$ includes 
label~1 points ---  then, for each $q \in
\partial R(n+1) \cap Q$, we  accept the chord $\{q, q' \}$ if $q' = q'(q)
\not\in R'(n+1)$ (that is, if the chord would reach beyond the cluster
$R'(n+1)$), we reject it if $q' \in R(n)$, 
and we accept the chord  $\{q, q' \}$ with probability $1/2$ if $q'
\in  \partial R(n+1) $, independently of all else.
Letting $Q(n+1) := \{q':\{q, q' \} \mbox{ newly accepted}\}$, 
take $R(n+1) = R'(n+1)\cup Q(n+1)$  and set $B(n+1) = B'(n+1) + 
|Q(n+1)|$. Note that~$B(n+1)$ may be either larger or smaller
than~$B(n)$, and that $B_{\lceil L/2 \rceil} = 1$~a.s.

After at most $\lceil L/2 \rceil$ time steps, each of the potential
chords has been either accepted or rejected independently with
probability~$1/2$, because of auxiliary randomization for those
chords such that $\{q,q'\} \in \partial R(n)$ for some~$n$, and
because of the random labelling of the end points of the chords 
for the remainder.  Hence this construction does indeed lead to
$\Po(L\r/2)$ independent uniform chords of~$C$.

For our analysis, as in~\cite{BR}, we define
a second process ${S}(n)$, starting from the same $P$
and the same set of potential chords, and with the same unit growth per
time step. The differences are that {\it every\/} potential
chord is included, so that no thinning takes place, and, additionally,
whenever two intervals intersect,
they continue to grow, overlapping one another, and each continues to
generate further chords according to
a Poisson process of rate~$\rho$. This 
pure growth process ${S}(n)$ agrees with the original construction
during
the initial development with high probability, until~$S$ has grown 
enough that overlap becomes likely;
its advantage is that it has a branching structure, and is thus much
more easily analysed. We denote its
length at time $n$ by $s(n)\geq r(n)$, overlaps now being counted
according to multiplicity, and the number of intervals by 
$M(n) \geq B(n)$. Then $M(n)$ is
just a pure birth chain  with offspring distribution $1+ \Po(2 \rho)$, 
so that 
  $\EE M(n) = (1 + 2 \rho)^n$, and the total length of the~$M(n)$ 
intervals is given by  
$$s(n) =  2 \sum_{k=0}^{n-1} M(k) ,$$
so that 
$$\EE s(n) = \rho^{-1} ((1 + 2 \rho)^n -1).$$
Furthermore,
$$W(n):=  (1 + 2 \rho)^{-n} M(n) $$
 forms a square integrable martingale, so that 
$ (1 + 2 \rho)^{-n} M(n) \rightarrow
W_\r$ a.s.\ for some $W_\r$ such that $W_\r>0$ 
a.s.\ and 
$\EE W_\r=1$.  Hence also $ (1 + 2 \rho)^{-n}s(n) \rightarrow
\rho^{-1}W_\r$ a.s.  and
$\frac{s(n)}{M(n)} \rightarrow \rho^{-1}$ a.s.. Note also that 
$ \var W(n)\leq 1.$

Our strategy is to pick a starting point $P$, and run both
constructions up to an integer time~$\t_r$, chosen in such
a way that $R(n)$ and~$S(n)$ are (almost) the same for
$n\le \t_r$. 
Pick
$$
n_0 = \left\lfloor \frac{\log\left({L\rho}\right) }
  {2\log(1+2 \rho)} \right\rfloor, 
$$ 
where $\lfloor x \rfloor$ denotes the largest integer 
no greater than~$x$, and let
$$ 
 \phi_0 := \f_0(L,\r) = (L\r)^{-1/2}(1+2\rho)^{n_0},
$$
so that $(1+2\rho)^{n_0} = \phi_0 \sqrt{{L\rho}}$
and $(1+2\r)^{-1} \le \f_0 \le 1$;
note that $\phi_0 \approx 1$ if~$\r$ is small.
Now let $\tau_r=n_0+r$, and assume that 
\begin{equation}\label{r-bound}
 \vert  r  \vert \leq \frac{1}{6\log(1 + 2 \rho)} \log \left({L \rho}\right) ,
\end{equation}  
implying in particular that  $\tau_r \leq 
\frac{ 2\log (L \rho) }{3\log(1 + 2 \rho)}$. 
Then, writing $R_r=R(\tau_r), { S}_r = {S}(\tau_r),
M_r = M(\tau_r)$, and $s_r = s(\tau_r)$, we have 
$$  \EE
M_r = \phi_0 \sqrt{L \rho} (1+ 2 \rho)^r  $$ 
and 
$$\EE s_r = \rho^{-1} (\phi_0\sqrt{ L \rho}   (1+ 2 \rho)^r -1).
$$

Next, independently and uniformly,  we pick a second
point $P' \in C$, and a second set of potential chords, $Q'$, and 
 run both
constructions for time $\tau_{r'}$, where~$r'$ also satisfies
\Ref{r-bound}, yielding $R'_{r'}, {S}'_{r'}, 
M'_{r'}=:N_{r'}$ and $s'_{r'}=:u_{r'}$.  Then, at least
for small~$\r$, there are about 
$   \phi_0^2 {L \rho} (1+ 2 \rho)^{r+r'}  $ 
pairs of intervals, with one in ${S}_r$ and the
other in ${S}'_{r'}$, and each is of typical length $\rho^{-1}$, so that
the expected number of intersecting
pairs of intervals  is about 
$$
\frac{2}{L \rho}  \phi_0^2 {L \rho} (1+ 2 \rho)^{r+r'} = 
2 \phi_0^2 (1+ 2 \rho)^{r+r'},
$$ 
which, in the chosen range of $r, r'$, grows
from almost nothing to the typically large value
 $2 \phi_0^2  (L\rho)^{1/3}$. For later use, label the 
intervals in $S_r$ as $I_1, \ldots, I_{M_r}$, and the intervals in
$S_{r'}'$ as $J_1, \ldots, J_{N_{r'}}$; then we can write
the number~$\whv_{r,r'}$ of intersecting pairs of intervals as
\bea \label{what}
\whv_{r,r'} = \sum_{i=1}^{M_r} \sum_{j=1}^{N_{r'}} X_{ij},
\ena
where
\bea  \label{xij}
X_{ij} = {\bf 1}\{ I_i \cap J_j \neq \emptyset\}. 
\ena

Now the probability
that $\whv_{r,r'}=0$ is the same as when the construction for ${S}'$
uses the original set~$Q$ of potential
chords, because of the independence of Poisson processes on disjoint
subsets;
the event $\whv_{r,r'}=0$ indicates that the two
processes have no intersecting pairs of intervals when
stopped at the times~$\tau_r$,
$\tau_{r'} $,  and thus use disjoint sets of chords. 
Furthermore, we can show that the
event $\whv_{r,r'}=0$ is with high probability the same as the event
$V_{r,r'}=0$, where $V_{r,r'}$ is the number of
intersections of $R(r)$ and $R'(r')$. Finally, if $R(r)$ and $R'(r')$ 
have no intersections, then the ``small worlds''
distance between $P$ and~$P'$ is more than 
$$
\tau_r+ \tau_{r'} = 2n_0 + r + r'.
$$
Hence we have solved the problem if we can find a good approximation
to the probability that $\whv_{r,r'}=0$; this we do by showing that
$\whv_{r,r'}$ approximately has a mixed Poisson distribution, and 
by identifying the mixture distribution. We usually take $r=r'$ or
$r=r'+1$, the latter to allow for the possibility of the number of 
steps in the shortest path being odd. 

After this preparation, we are in a position to summarize our
main results. These are treated in more detail in the next section, 
in Theorem \ref{corollary6}, Corollary \ref{rholarge} and
Theorem~\ref{corollary14}.
We let~$D$ denote the small worlds distance between a randomly
chosen pair of points $P$ and~$Q$ on~$C$, so that, as above,
$$
\PP[D > 2n_0 + r + r'] = \PP[V_{r,r'} = 0].
$$
The following theorem approximates the distribution of~$D$ by that
of another random variable~$D^*$, whose distribution is more
accessible;
in this theorem, $\rho$ and the derived quantities $\phi_0, n_0, N_0$ 
and~$x_0$ all implicitly depend on~$L$, as does the distribution of~$D^*$.

\begin{theorem} \label{summary}
Let $\Delta$ denote a random variable on the integers with distribution 
given by
\beas
\PP\left[\Delta > x \right]
= \EE \{  e^{-2 \phi_0^2(1+2\rho)^{x} W_\r W'_\r} \},
  \quad x \in \ZZ,
\enas
and set $D^* = \Delta + 2n_0  $.
If $L\r \to \infty$ and $\r = \r(L) = O(L^{\b})$,
with $\b < 4/31$, then 
\beas
{\dtv({\cal L}(D), {\cal L}(D^*)) }\rightarrow 0 \quad\mbox{as}\quad
L \rightarrow 
\infty.
 \enas  
\begin{enumerate} 
\item If $\r$ is large, let $N_0$ be such that 
$(1 + 2 \r)^{N_0} \le L \r < (1 + 2 \r)^{N_0+1}$, and define 
$\alpha \in [0,1)$ to be such that $L \r = (1 + 2 \r)^{N_0+\alpha}$; 
then, with
$x_0=N_0 - 2 n_0 + 1$, 
\beas
\PP \left[\Delta \ge x_0 \right] &\ge& 1 - 2 (1 + 2 \r)^{-\alpha};\\
\PP \left[\Delta \ge x_0 +1\right]
    &=&O\left( (1 + 2\r)^{-1+\alpha} \log(1+\r)\right),
\enas 
so that $\Delta$ concentrates almost all its mass on~$x_0$, unless $\alpha$ 
is very close to~$1$. 
\item If $\rho \rightarrow 0$, the distribution of $\rho \Delta$ approaches 
that of the random variable~$T$ defined in \cite{BR}, Corollary~3.10:
\beas
\PP[\r\D > x] \to
\PP \left[T > x\right] = \int_0^\infty \frac{e^{-y}}{1 + 2 e^{2x}y} \,dy.
\enas  
\end{enumerate} 
\end{theorem} 

\noindent The errors in these  distributional approximations
 are also quantified, for given choices of $L$ and $\rho(L)$.   

\ignore{
To understand this result better, it is instructive to calculate the 
Laplace transform of the limiting variable $T$. We have that 
\beas
\phi_T(t) &=& \EE e^{tT}\\
&=& \EE \int_{-\infty}^T t e^{ty} dy \\
&=& t \int_{-\infty}^{\infty} \PP(T > y) \, dy\\
&=& t  \int_{-\infty}^{\infty}\int_0^\infty e^{tx} 
\frac{e^{-y}}{1 + 2 e^{2x} y} \, dy \, dx.  
\enas 
Using the substitution $u=2x$ and then $v=u+\log (2y)$ we obtain
\beas
\phi_T(t) &=& \frac{t}{2}   \int_{-\infty}^{\infty}\int_0^\infty 
e^{\frac{t}{2} v} (2y)^{-\frac{t}{2}} \frac{e^{-y}}{1 + e^v} \, dv \, dy\\
&=& \frac{t}{2} 2^{-\frac{t}{2}} \Gamma\left(1 - \frac{t}{2}\right) 
\int_0^\infty  \frac{e^{-y}}{1 + e^v} \, dv  .
\enas 
Substituting $z = (1+e^v)^{-1}$ we recognize the last integral 
as a Beta integral, giving 
\beas
\phi_T(t) &=&  \frac{t}{2} 2^{-\frac{t}{2}} 
\Gamma\left(1 - \frac{t}{2}\right) 
B\left( \frac{t}{2}, 1 - \frac{t}{2}\right)\\
&=&  \frac{t}{2} 2^{-\frac{t}{2}} \Gamma\left(1 - \frac{t}{2}\right)^2
\Gamma\left(\frac{t}{2}\right)  \\
&=&  2^{-\frac{t}{2}} \Gamma\left(1 - \frac{t}{2}\right)^2
\Gamma\left(1 + \frac{t}{2}\right) .
\enas 
Thus we can write $T$ as the sum of three independent random variables.  
Let $Z_1, Z_2, Z_3$ be i.i.d. $exp(1)$-variables, and put
\beas
Y_1 &=& 2 \log Z_1\\
Y_2 &=&  2 \log Z_2\\
Y_3 &=& - 2 \log Z_3.  
\enas 
Then 
\beas
T \stackrel{d}{=} Y_1 + Y_2 + Y_3 - \log 2. 
\enas 
Here, the variables $Z_2$ and $Z_2$ correspond to $W$ and $W'$; $-\log 2$ 
gives the factor 2. 
The third variable, $Z_3$, corresponds to the waiting time to the 
first event
in  a Poisson process with rate 1. The rate 1 indicates a new, 
random time scale:  Given $W, W'$, rescale
$t \rightarrow 2 e^{2t} W W'$. Then the waiting time until the 
first encounter of the two independent processes in the approximating 
continuous model is exactly $exp(1)$-distributed. 
In contrast, similar calculations reveal that for the approximating 
random variable $T_{NMW}$ in \cite{NewmanWatts2} we would obtain
\beas
T_{NMW} \stackrel{d}{=} Y_1 + Y_2. 
\enas 
The discrepancy between the NMW heuristic and the true asymptotics 
hence stems from having ignored the random nature of the waiting time 
until the two processes first meet, on an exponentially transformed 
random time scale. 
This also explains the different asymptotic behaviour in the 
discrete-continuous
case. As the time to first encounter is now discrete, we integrate 
an exponential variable over an interval of size one in the new, 
exponential time scale. Depending on the parameters this interval 
may already capture almost all the mass of that exponential variable. 
Moreover, using the Laplace transform we can obtain moments, and 
cumulants $\kappa_\ell$. 
Differentiating the cumulant generating function 
$\psi_T(t) = \log \phi_T(t)$ 
gives ($\gamma$ being Euler's constant, and $\zeta$ being the 
Riemann zeta function) 
\beas
\EE T &= &\frac12(\gamma - \log 2) \approx 0.058\\
\var T &=& \frac{{\pi}^2}{6}\\
\kappa_\ell &=& \left\{ \begin{array} {l l } 
3 \times 2^{-\ell} (\ell-1)! \zeta(\ell) & \quad \ell \ge 2 \mbox{ even } \\
 2^{-\ell} (\ell-1)! \zeta(\ell) & \quad \ell \ge 3 \mbox{ odd. } 
\end{array}
\right. 
\enas 
In comparison
\beas
\EE T_{NMW}= 0, \quad \var T_{NMW} = \frac{{\pi^2}}{12}. 
\enas 
Formally, we can also calculate the Laplace transform for 
$\Delta = \Delta_\r$;
\beas
\phi_{\Delta_\r}&=& (2 \phi_0^2)^{-t} \Gamma(1+t) 
\left( \EE W_\r^{-\frac{t}{\log (1+2 \r)}} \right)^2,
\enas 
so that
\beas
\EE \log(1 + 2 \r) \Delta_\r&=& -\gamma - \log 2 - 2 \log \phi_0 - 2 
\EE (\log W_\r).  
\enas 
Unfortunately in general $ \EE \log W_\r$ does not appear to be known 
(although $W_\r$ is known to have a continuous density, see Theorem 
6.1, p.89., in \cite{Asmussenhering}).
}

This result shows that, for~$\r$ small and $x=l\r$ with $l\in\ZZ$,
\bea
\PP[\r\D > x] &=& \EE\{e^{-2\f_0^2(1+2\r)^{x/\r}W_\r W'_\r}\}\non\\
  &\approx& \EE\{e^{-2e^{2x}WW'}\} = \PP[T>x], \label{2-exp}
\ena
where $W$ and~$W'$ are independent NE$(1)$ random variables.
Indeed, it follows from Lemma~\ref{phi-phie} below
that $W_\r\to_{{\cal D}}W$
as $\r\to0$.  One way of realizing a random variable~$T$ with
the above distribution is to realize $W$ and~$W'$, and then to
sample~$T$ from the conditional distribution 
\bea
\PP[T>x \giv W,W'] &=& e^{-2e^{2x}WW'} \non\\
  &=& e^{-\exp\{2x+\log2+\log W+\log W'\}}\non\\
  &=& e^{-\exp\{2x+\log2 - G_1 - G_2\}},\label{2-gum}
\ena
where $G_1 := -\log W$ and $G_2 := -\log W'$ both have the
Gumbel distribution.  With this construction,
\beas
\PP[2T - \{G_1+G_2-\log2\} > x \giv W,W'] = e^{-e^x},
\enas
whatever the values of $W$ and~$W'$, and hence of $G_1$ and~$G_2$,
implying that
$$
2T =_{{\cal D}} G_1 + G_2 - G_3 - \log2,
$$
where $G_1,G_2$ and $G_3$ are independent random variables with
the Gumbel distribution.  The cumulants of~$T$ can thus immediately
be deduced from those of the Gumbel distribution, given in
Gumbel~\cite{Gum}:
\beas
\EE T &= &\frac12(\gamma - \log 2) \approx -0.058;\\
\var T &=& \frac{{\pi}^2}{8}.
\enas

Note that, in view of Corollary~\ref{cor2} below, the conditional
construction~\Ref{2-gum} can be interpreted in terms of the
processes $S$ and~$S'$, since $W_\r$ and~$W'_\r$ are essentially
determined by the early stages of the respective pure birth
processes, and the extra randomness, conditional on the values
of $W_\r$ and~$W'_\r$, comes from the random arrangement of the
intervals on the circle~$C$.

In the NMW heuristic, the random variable~$T_{NMW}$ is logistic,
having distribution function $e^{2x}(1+e^{2x})^{-1}$; note that
this is just the distribution of $\half(G_1-G_3)$.  Hence the
heuristic effectively neglects some of the initial branching
variation.

\section{The continuous circle model: proofs}
 \setcounter{equation}{0}

The first step in the argument outlined above is to establish a Poisson
approximation theorem for the number of pairs of overlapping intervals,
one in~$S_r$ and the other in~$S'_{r'}$.  The following
 result has been shown in \cite{BR}.

\begin{proposition}\label{cor1}
Let $M$ intervals $I_1,\ldots,I_M$ with lengths $t_1,\ldots,t_M$ and
$N$ intervals $J_1,\ldots,J_N$ with lengths $u_1,\ldots,u_N$ be
positioned
uniformly and independently on~$C$.  Set $V:= \siiM\sjN \Xij$, where
$\Xij := I[I_i\cap J_j \neq \emptyset]$. Then
$$
\dtv(\law(V),\Po(\l_{(M,N,t,u)})) \le 4(M+N)v_{tu}/L,
$$
where $\l_{(M,N,t,u)} := L^{-1}(Nt+Mu)$, $t:=\siiM t_i$,
$u:=\sjN u_j$ and $v_{tu} := \max\{\max_i t_i,\max_j u_j\}$.
\end{proposition}

\bn
The proposition translates immediately into a useful statement about
$\whv_{r,r'}$,
when~$P'$ is chosen uniformly at random, independently of all else.

\begin{corollary}\label{cor2}
For the processes $S$ and~$S'$ of the previous section, we have
\beas
&&|\PP[\whv_{r,r'} = 0\giv M_r=M,N_{r'}=N,s_r=t,u_{r'}=u] -
\exp\{-L^{-1}(Nt+Mu)\}|\\
&&\qquad \le 8L^{-1}(M \tau_r +N \tau_{r'}).
\enas
\end{corollary}

\remark  If~$P'$ is not chosen at random, but is a fixed point of~$C$,
the result of Corollary~\ref{cor2} remains essentially unchanged,
provided that $P$ and~$P'$ are more than an arc distance of~$\tau_r
+\t_{r'}$ apart.  The only difference is that then $X_{11}=0$ a.s., and that
$Nt+Mu$ is replaced by $Nt+Mu - 2\tau_r - 2 \tau_{r'}$.  If $P$ and~$P'$ are
less than~$\tau_r+ \tau_{r'} $ apart, then $\PP[\whv_{r,r'}=0] = 0$.

\medskip

The next step is to show that $\PP[\whv_{r,r'}=0]$ is close to 
$\PP[V_{r,r'}=0]$.
We do this by directly comparing the random variables 
$\whv_{r,r'}$ and~$V_{r,r'}$ 
in the joint construction. As for Corollary~3.5 in~\cite{BR}, 
the following assertion can easily 
be shown to hold.

\begin{proposition}\label{cor3}
With notation as above, we have
$$
\PP[\whv_{r,r'}\neq V_{r,r'}] \le 32\tau_r \tau_{r'} L^{-2}\EE\{\tfrac12
M_rN_{r'}(M_r+N_{r'}-2)\}.
$$
\end{proposition} 

To apply Corollary \ref{cor2} and Proposition ~\ref{cor3}, it remains 
to establish
more detailed information about the distributions of $M_r$ and~$s_r$.
In particular, we need to bound the first and second moments of~$M_r$,
and to approximate the quantity $\EE(\exp\{-L^{-1}(N_{r'}s_r+M_ru_{r'})\})$.
We begin with the following lemma.

\begin{lemma} \label{Lemma1}\label{2.01}
The random variable $M(n)$ has as probability generating function
\beas
G_{M(n)}(s) := \EE s^{M(n)} = f^{(n)}(s), \hh f(s) = se^{2 \rho (s-1)},
\enas 
where $ f^{(n)}$ denotes the $n$th iteration of $f$. 
In particular,  we have  
\beas
 \EE M_r &=& \phi_0 {\sqrt{L \rho}} (1+ 2 \rho)^r  
\nonumber \\  
\frac{1}{2} \EE M_r(M_r - 1) 
&=& (\rho + 1) \phi_0 {\sqrt{L \rho}} (1+ 2 \rho)^{r-1}
\left\{   \phi_0 {\sqrt{L \rho}} (1+ 2 \rho)^r -1
\right\} \nonumber\\ 
&\leq& \phi_0^2 {L \rho} 
(1+ 2 \rho)^{2r}.  \enas
\end{lemma}

\proof 
Since $M(n)$ is a branching process with $1 + \Po(2\rho)$ 
offspring distribution, the probability generating function is
immediate, as are the moment calculations 
\beas \label{2.4}
\EE M(n) &=& (1 + 2 \rho)^n;\\ 
\EE M(n)(M(n)-1) &=& 2 (\rho + 1) (1 + 2 \rho)^{n-1} 
     \left\{(1 + 2 \rho)^n -1\right\} .
\enas  
The moments of~$M_r$ follow from the definition of $\tau_r$. 
\ep

These estimates can be directly applied in Corollary \ref{cor2}
and Proposition~\ref{cor3}. Define
\bea 
\h_1(r,r') &:=& 64 \{\r( n_0+(r \vee r'))\} ^2 \phi_0^3 
    (1+2\rho)^{r+r'+(r\vee r')} \label{eta1} \\
\h_2(r,r') &:=& 16\{\r(n_0 + (r\vee r'))\} \phi_0  
    (1+ 2\rho)^{(r\vee r')}. \label{eta2} 
\ena

\begin{corollary}\label{cor4}
We have 
$$
\PP[\whv_{r,r'}\neq V_{r,r'}] \le \h_1(r,r') (L\r)^{-1/2}
$$ 
and 
$$
|\PP[V_{r,r'} = 0] - \EE\exp\{-L^{-1}(N_{r'}s_r+M_ru_{r'})\}| \le
   \{\h_1(r,r') + \h_2(r,r')\}(L\r)^{-1/2}.
$$
\end{corollary}

\noindent Consideration of the quantity
$\EE(\exp\{-L^{-1}(N_{r'}s_r+M_ru_{r'})\})$ now gives
the  immediate asymptotics of
$$
\PP[V_{r,r'} = 0] = \PP[D > 2n_0 + r + r'],
$$
where~$D$ denotes the ``small world'' distance between $P$ and~$P'$. 

\begin{corollary}\label{cor5}
If $\r = \r(L)$ is bounded above and $L\r \to \infty$, then 
as $L \to \infty$,
\beas
&&|\PP[D > 2n_0 + r + r'] - \EE\exp\{-2 \phi_0^2 (1 + 2 \rho)^{r+r'} 
   W_\r
W'_\r\}| \rightarrow 0
 \enas
uniformly in 
$|r|, |r|' \le \frac{1}{6\log(1 + 2 \rho)} \log \left({L \rho}\right)$,
where $W_\r$ and $W'_\r$ are independent copies of 
the limiting random variable associated with the pure birth
chain~$M$.  
\end{corollary}

\proof
The conditions ensure that $\t_r$ and~$\t_{r'}$ both tend to
infinity as $L\to\infty$, at least as fast at $c\log(L\r)$, for
some $c>0$.  Then, 
since $W(n)=(1 + 2 \rho)^{-n} M(n) \to W_\r$
a.s.\ and $s(n)/M(n) \to \rho^{-1}$ a.s., and since $(1 + 2
\rho)^{\tau_r+ \tau_{r'}} = \phi_0^2 (1 + 2 \rho)^{r+r'}{L\rho}
$, it is clear that
\beas 
\exp\{-L^{-1}(N_{r'}s_r+M_ru_{r'})\} &\sim &\exp\{-2(L\rho)^{-1} M_r
N_{r'} \}\\ 
&=& 
\exp\left\{ - 2 (L\rho)^{-1}  (1 + 2 \rho)^{ \tau_r+ \tau_{r'} }
  W(\t_r) W'(\t_{r'})\right\}\\ 
&\sim& \exp\{-2
\phi_0^2 (1 + 2 \rho)^{r + r'}  W_\r {W'_\r}\},
\enas 
uniformly for $r$, $r'$ in the given ranges. \hfil$\Box$

Hence $\PP[D > 2n_0 + r + r']$ can be approximated in terms of the 
distribution of
the limiting random variable~$W_\r$ associated with the  pure birth
chain~$M$.  However, in contrast to the model with time running
continuously, this distribution is not always $\NE(1)$, but
genuinely depends on~$\rho$.  Its properties are not so easy to
derive, though moments can be calculated, and, in particular,
\begin{equation}\label{w-moments}
\EE W_\r = 1;\qquad \var W_\r = 1/(1+2\r);
\en
it is also shown in Lemma~\ref{phi-phie} that $\law(W_\r)$ is
close to~$\NE(1)$ for~$\r$ small.  We also need the following
lemma, which is useful in bounding the behaviour of the upper 
tail of~$\law(D)$.

\begin{lemma}\label{D-tail}
For all $\th,\r > 0$,
$$
\EE(e^{-\th W_\r W'_{{\rho}}}) \le \th^{-1}\log(1+\th).
$$
\end{lemma}
 
\proof
The offspring generating function of the birth process~$M$ satisfies
$$
f(s) = se^{2\r(s-1)} \le s\{1+2\r(1-s)\}^{-1} =: f_1(s)
$$
for all $0\le s\le 1$.  Hence, with $m=1+2\r$,
\begin{equation}\label{f-lim}
\EE(e^{-\ps W_\r}) = \lnti f\un(e^{-\ps m^{-n}})
  \le \lnti f_1\un(e^{-\ps m^{-n}}) = (1+\ps)^{-1}.
\en
The last equality follows from (8.11), p.17 in \cite{Harris}, noting that the 
right-hand side is the Laplace transform of the NE(1) - distribution.  
Furthermore, we have
$$
(1+\th w)^{-1} = \th^{-1} \int_0^\infty e^{-tw} e^{-t/\th}\,dt,
$$
and so, applying~\Ref{f-lim} twice, and because the function
$(1+t)^{-1}$ is decreasing in $t\ge0$, we obtain
\beas
\EE\Bl e^{-\theta W_\r W'_{{\rho}}}\Br &\le& \EE\{(1 + \th W_\r)^{-1}\}\\
&=& \theta^{-1} \int_0^{\infty} \EE e^{-t W_\r } e^{-t/\th} \, dt\\  
  &\le& \th^{-1}\int_0^\infty (1+t)^{-1} e^{-t/\th}\,dt \\
  &\le& \th^{-1}\int_0^\th(1+t)^{-1}\,dt = \th^{-1}\log(1+\th),
\enas
as required.  \hfil$\Box$

The simple asymptotics of Corollary~\ref{cor5} can be sharpened.
At first sight surprisingly, it turns out that it is not 
necessary for the times $\t_r$
and~$\t_{r'}$ to tend to infinity, since, for values of~$\r$ so
large that~$n_0$ is bounded, the quantities~$W(n)$ are (almost)
constant for all~$n$. Write 
\bea \label{sdecomp}
(1+2\rho)^{-n}s(n)&=& 2  \sum_{j=0}^{n-1}W(j) (1+2\rho)^{-(n-j)}\nonumber \\
&=&  \frac{1}{\rho}\, W(n) + \frac{1+\rho}{\rho}\, U(n),
\ena 
where
\beas
\frac{1+\rho}{\rho} \,U(n) &=& 2 \sum_{j=0}^{n-1} (W(j)-W(n)) (1+2\rho)^{-(n-j)} -
W(n) \rho^{-1}(1+2\rho)^{-n}.
\enas
Computation gives $\EE U(n)=- (1+\rho)^{-1}(1+2\rho)^{-n}$, and
$$
\EE \{ (W(n)-W(j))(W(n)-W(\ell)) \} 
= \frac{1}{(1+2\rho)^{j+1}}\left( 1 - \frac{1}{(1+2\rho)^{n-j}}
\right)
$$
if $j \geq \ell$, so that
\bea \label{uvar}
\var\{U(n)\} &\leq& 2 \,\frac{2(1+2\rho)^{-n} + 2 (1+2  \rho)^{-2n}
}{(1+\rho)^2}\nonumber \\ &\leq& 8\, \frac{(1+2\rho)^{-n} }{(1+\rho)^2},
\ena
and thus 
\begin{equation}\label{star-3}
(1 + \r)^{2} \EE\{ U(n)^2\} \le 9(1+2\r)^{-n}.
\en
Then we have
\beas
\lefteqn{L^{-1}(N_{r'}s_r+M_ru_{r'})}\\ &=&
 \phi_0^2(1+2\rho)^{r+r'} \{W(\tau_r)(W'(\tau_{r'})+(1 + \rho) U_{r'}') + W'(\tau_{r'})(W(\tau_r)+(1 + \rho) U_r)\} ,
\enas
where $W(\tau_r) := W(\t_r)$ and $U_r := U(\t_r)$,
so that, by Taylor's expansion, and because $\EE W(n) = 1$ for all~$n$,
\bea
&&\left| \EE \exp\{- L^{-1}(N_{r'}s_r+M_ru_{r'})\}  - 
\EE  \exp\{-2 \phi_0^2(1+2\rho)^{r+r'}  W(\tau_r) W'(\tau_{r'})\}
    \right| \non \\ 
&&\qquad\leq \phi_0^2(1+2\rho)^{r+r'}(1 + \rho)  \{\EE |U_r| + \EE |U'_{r'}| \}  
   \label{star-1}
\ena
and
\bea
&&\left| 
\EE \exp\{-2 \phi_0^2(1+2\rho)^{r+r'} W(\tau_r) W'(\tau_{r'})\}
  -  \EE\exp\{-2 \phi_0^2(1+2\rho)^{r+r'} W_\r W'_\r\}    
   \right| \non\\
&&\qquad\leq 2\phi_0^2(1+2\rho)^{r+r'} \{\EE | W_\r-W(\tau_r) | 
 + \EE | W'_\r-W'(\tau_{r'})|\}. \label{star-2}
\ena
Using these results, we obtain the following theorem.

\begin{theorem}\label{D4}
If $P'$ is randomly chosen on~$C$, then 
\beas
&&\left| \PP[D>2n_0+r+r'] 
  -\EE \{  e^{-2 \phi_0^2(1+2\rho)^{r+ r'} W_\r W'_\r}
      \}\right|  \\
&&\qquad\leq \{\h_1(r,r') + \h_2(r,r')\}(L\r)^{-1/2} 
       + \h_3(r,r')(L\r)^{-1/4},
  \enas 
where $\h_1, \h_2$ are given in (\ref{eta1}) and (\ref{eta2}), 
$$
\h_3(r,r') := 
   10 \phi_0^{3/2}(1+2\rho)^{r+r'- \frac12(r \wedge r') }
$$   
and where, as before, $D$ denotes the shortest distance
between $P$ and~$P'$ on the shortcut graph. 
\end{theorem}

\proof  Since $\{V_{r,r'}=0\} = \left\{ D > 2n_0+r + r' \right\}$,
we use Corollary~\ref{cor4} and \Ref{star-1} and~\Ref{star-2}
to give
\bea
\lefteqn{\left| \PP[D>2n_0+r+r'] 
   -\EE \{  e^{-2 \phi_0^2(1+2\rho)^{r+ r'} 
        W_\r {W_\r}'} \}\right| } \non\\
&\leq& \{\h_1(r,r') + \h_2(r,r')\}(L\r)^{-1/2} + \phi_0^2(1+2\rho)^{r+r'} 
(1 + \rho)  \{\EE | U_r| + \EE | U_{r'}' | \} \non\\
 &&\quad \mbox{} + 2\phi_0^2(1+2\rho)^{r+r'} \{\EE | W_\r-W(\tau_r) | 
        + \EE | W'_\r-W'(\tau_{r'})|\}.\label{eta-3}
\ena
Now, from~\Ref{star-3} and the Cauchy-Schwarz inequality,
\begin{equation}\label{term-1}
(1 + \r) \EE|U_r| \le 3(L\r)^{-1/4}\f_0^{-1/2}(1+2\r)^{-r/2}.
\en
Then, since $W(n)$ is a martingale, and       
\beas
W_\r- W(n) &=& \sum_{\ell=n}^\infty (W(\ell + 1) - W(\ell))\\
&=& \sum_{\ell=n}^\infty (1 + 2 \rho)^{- \ell-1} (M(\ell+1) - (1 + 2 \rho)
M(\ell)) ,
\enas
we have 
\beas
\EE (W_\r- W(n))^2 &=& \sum_{\ell=n}^\infty 
  (1 + 2 \rho)^{-2(\ell+1)} \EE (M(\ell+1) - (1 + 2 \rho)M(\ell))^2.
\enas  
Now
\beas
M(\ell+1) 
&=& \sum_{i=1}^{M(\ell)}(Z_{\ell+1}(i) +1 ),
\enas 
where $(Z_\ell(i))_{\ell, i}$ are i.i.d.\ $\Po(2 \rho)$-variates,
and so 
\beas
\EE   (M(\ell+1) - (1 + 2 \rho)M(\ell))^2
&=& \EE \var(M(\ell+1)\giv M(\ell))\\ 
       &=& 2\r\EE M(\ell) = 2 \rho (1 + 2 \rho)^\ell, 
\enas 
implying that
\beas
\EE (W_\r-W(\tau_r))^2 &\leq& 
 2 \rho (1 + 2 \rho)^{-2} \sum_{\ell=\t_r}^\infty (1 + 2 \rho)^{-\ell} \\
&=&  (1 + 2 \rho)^{-2}(1 + 2 \rho)^{- \t_r}.
\enas
Hence
\beas
\EE | W_\r-W(\tau_r) | &\leq& 
  (1 + 2 \rho)^{-1 - \frac12 (n_0 + r)} \\
&\le&  (L\r)^{-1/4}\f_0^{-1/2}(1 + 2 \rho)^{-\frac{r}{2}},
 \enas 
and the theorem follows.  
\ep

Theorem~\ref{D4} can be translated into a uniform distributional 
approximation, as follows.

\begin{theorem}\label{corollary6} 
If $\Delta$ denotes a random variable on the integers with 
distribution given by
\begin{equation}\label{T-def}
\PP\left[\Delta > x \right]
= \EE \{  e^{-2 \phi_0^2(1+2\rho)^{x} W_\r {W_\r}'} \},
  \quad x \in \ZZ,
\en
and $D^* = \D + 2n_0$, then
\beas
\lefteqn{\dtv({\cal L}(D), {\cal L}(D^*)) }\\
&=& O\left( \log(L\r)(1 + 2 \r)^{1/4} (L \rho)^{-\frac{1}{7}} +  
    \left(\frac{\r\log(L \rho)}{\log(1+2\rho)}\right)^2 
      (1 + 2 \r)^{1/2}(L\r)^{-\frac27}\right). 
 \enas  
\end{theorem}
 
In particular, for $\r = \r(L) = O(L^{\b})$ with $\b <4/31$,
\beas
{\dtv({\cal L}(D), {\cal L}(D^*)) } \rightarrow 0 \quad 
\mbox{as}\quad L \rightarrow \infty.
\enas

\proof
It is easy to see that $\Delta$, defined as above, is indeed a 
random variable. 
Its upper tail is bounded by Lemma~\ref{D-tail}, which implies that
\begin{equation}
\label{T-tail}
\PP[\Delta >x] \leq \frac1{2\phi_0^2} (1+2\r)^{-x}(2 + x\log(1+2\r))
\end{equation}
for any $x>0$, since $\phi_0 \le 1$ and
$\log(1+2y) \le 2 + \log y$ in $y\ge1$. Then, for any~$x\in\ZZ$, 
writing $r(x)=\lfloor x/2 \rfloor$ and $r'(x)=x-r(x) \le (x+1)/2$, 
it follows
from Theorem~\ref{D4} that
\beas
&&|\PP[D > 2n_0 + x] - \PP[D^* > 2n_0 + x]|\\ 
&&\qquad \le \{\h_1(r(x),r'(x)) + \h_2(r(x),r'(x))\}(L\r)^{-1/2}
     + \h_3(r(x),r'(x))(L\r)^{-1/4}\\
&&\qquad = O\Bl 
 \left( \frac{\r \log(L \rho)}{\log(1+2\rho)} 
 \right)^2  (1 + 2 \r)^\frac12 (L\r)^{-\frac{2}{7}} \right.\\
&&\left.\qquad\qquad\qquad 
  \mbox{} +  \left( \frac{\r \log(L \rho)}{\log(1+2\rho)} \right)  
    (1 + 2\r)^\frac12 (L\r)^{-\frac37}
     +(1 + 2 \rho)^\frac14 (L \rho)^{-\frac{1}{7}} 
      \vphantom{\left(\frac{\r \log(L \rho)}{\log(1+2\rho)}\right)^2}\Br,
\enas
so long as 
$x \le \lfloor \frac{\frac17\log(L\r) - 2 \log \phi_0}{\log(1+2\r)}\rfloor$. 
This is combined with~\Ref{T-tail} evaluated at 
$x = \lceil \frac{\frac17\log(L\r) - 2 \log \phi_0}{\log(1+2\r)}\rceil$,
which gives rise to a term of order
$O \left( (L\r)^{-\frac17} \log(L\r) \right)$, and the main estimate
follows.
 
The above bound tends to zero  as $L \rightarrow \infty$ as long as
$\r = \r(L) = O(L^{\b})$ for $\b < 4/31$. 
Thus the theorem
is proved.     
\ep

 For larger~$\r$ and for~$L$ large, it is
easy to check that~$n_0$ can be no larger than~$4$, 
 so that interpoint
distances are extremely short, few steps in each branching process
are needed, and the closeness of $\law(D)$ and~$\law(D^*)$ could
be justified by direct arguments.  Even in the range covered by
Theorem~\ref{corollary6}, it is clear that~$\law(D)$ becomes
concentrated on very few values, once~$\r$ is large, since the
factor $2\f_0^2(1+2\r)^x$ in the exponent in~\Ref{T-def} is
multiplied by the large factor $(1+2\r)$ if~$x$ is increased
by~$1$.  The following corollary makes this more precise. 

\begin{corollary} \label{rholarge} If~$N_0$ is such that
$$
(1+2\r)^{N_0} \le L\r < (1+2\r)^{N_0+1},
$$
and if $L\r = (1+2\r)^{N_0+\a}$, for some $\alpha \in [0,1)$, then, taking $x_0 = N_0 - 2n_0 + 1$,
we have
$$
\PP[\Delta \ge x_0]  \ge 1 - 2(1+2\r)^{-\a},
$$
and
 $$
\PP[\Delta \ge x_0 +1] = \EE\exp\{-2(1+2\r)^{1-\a}W_\r W'_{{\rho}}\}
 \le \half(1+2\r)^{-1+\a}\log(3+4\r).
$$   
\end{corollary}

\proof
The result follows immediately from Jensen's inequality; 
\beas
\EE\exp\{-2(1+2\r)^{-\a}W_\r W'_{{\rho}}\}
  & \ge& \exp \left\{ -2(1+2\r)^{-\a} \EE W_\r \EE W'_{{\rho}}  \right\} \\
&\ge&  1 - 2(1+2\r)^{-\a}
\enas
as $\EE W_\r = 1$, and from Lemma~\ref{D-tail} with
$\th = 2(1+2\r)^{1-\a}$. 
\ep

\nin Thus the distribution is essentially concentrated on the single
value~$x_0$ if~$\r$ is large and~$\a$ is bounded away from $0$ and~$1$.  
If, for instance,
$\a$ is close to~$1$, then both $x_0$ and $x_0+1$ may carry appreciable
probability.

If $\r\to\r_0$ as $L\to\infty$, then the distribution of~$\D$ becomes
spread out over~$\ZZ$, converging to a non--trivial limit as
$L\to\infty$ along any subsequence such that $\f_0(L,\r)$ converges.
Both this behaviour and that for larger~$\r$ are quite different
from the behaviour found in the continuous model
of~\cite{BR}.  However, if~$\r$ becomes smaller, the differences
become less; we now show that, as $\r\to0$, the distribution
of~$\r \Delta$ approaches the limiting distribution of $T$ obtained 
in~\cite{BR}.

The argument is based on showing that the distribution of $W_\r$
is close to~$\NE(1)$.  To do so, we employ the  
characterizing Poincar\'{e} equation for Galton--Watson branching 
processes (see Harris~\cite{Harris}, Theorem~8.2, p.15); if 
$$
\phi_\r(\theta) = \EE e^{- \theta W_\r}
$$
is the Laplace transform of ${\cal L}( W_\r)$, then
\bea \label{chareq}
 \phi_\r((1 + 2 \rho)\theta) = f(\phi_\r(\theta)).
\ena
We show that when $\rho \approx 0$ then $\phi_\r(\th)$ is close to
$\phi_e(\theta)  =  (1 + \theta)^{-1}$, 
the Laplace transform of the~$\NE(1)$ distribution.
 
Let 
$$ {\cal G} = \left\{g: [0, \infty) \rightarrow \RR:\, \|g \|_{\cal G} 
:=  \sup_{\theta > 0} \th^{-2}| g(\theta)| < \infty\right\},$$
and let
$${\cal H} = \left\{ \chi: [0, \infty) \rightarrow \RR: \chi (\theta) = 1 - \theta
+ g(\theta) \mbox{ for some }g \in {\cal G}\right\}.$$
Then ${\cal H}$ contains all Laplace transforms of probability distributions
with mean~$1$ and finite variance. 
On ${\cal H}$, define the operator $\oper$ by 
$$
(\oper \chi)(\theta) = f\left( \chi\left( \frac{\theta}{m} \right) \right),
$$
where
$$f(s) = s e^{2 \rho (s-1)}$$
is the probability generating function of $1 + \Po (2 \rho)$, and $m=1+2
\rho > 1$. Thus the Laplace transform~$\phi_\r$ of interest to us is a fixed
point of~$\oper$.

\begin{lemma} \label{cont}
The operator $\oper$ is a contraction, and, 
for all $\chi, \psi \in {\cal H}$, 
\beas
\| \oper \chi - \oper \psi \|_{\cal G} 
  &\leq& \frac{1}{m} \,\| \chi - \psi
\|_{\cal G}. 
 \enas
\end{lemma}

\proof
For all  $\chi, \psi \in {\cal H}$ and $\theta > 0$, we have
\beas
\th^{-2}| \oper \chi(\theta) - \oper \psi (\theta) | &=& 
  \th^{-2}\Blm f\left( \chi\left( \frac{\theta}{m} \right) \right) 
    - f\left( \psi\left( \frac{\theta}{m} \right) \right)\Brm  \\
&\leq& \| f \|_{\infty} \,\th^{-2}\Blm\chi\left( \frac{\theta}{m}
  \right) - \psi\left( \frac{\theta}{m} \right)\Brm \\
&=& \th^{-2}m \Blm\chi\left( \frac{\theta}{m}\right) -
   \psi\left( \frac{\theta}{m} \right)\Brm \\
&=& m^{-1}(\th/m)^{-2} \Blm\chi\left( \frac{\theta}{m}\right) -
   \psi\left( \frac{\theta}{m} \right) \Brm\\
&\leq&  m^{-1}\| \chi - \psi\|_{\cal G},
\enas
as required. 
\ep

\begin{lemma} \label{phie}
For the Laplace transform $\phi_e$, we have
\beas
\| \oper \phi_e  - \phi_e\|_{\cal G}  
   \leq \frac{2 \rho^2}{(1 + 2 \rho)^2}.
 \enas
 \end{lemma}

\proof
For all $\theta>0$, we have 
\beas
\left|\frac{| \oper \phi_e (\theta) - \phi_e(\theta)
}{\theta^2}\right| &=& \frac{1}{1 + \theta}\frac{1}{\theta^2} \left|
\left( 1 + \frac{2 \rho \theta}{m+  \theta} \right)e^{- 2 \frac{\rho
\theta}{m+ \theta}} - 1 \right| \\
&\leq&  \frac{1}{2(1 + \theta)\theta^2}\left( \frac{2 \rho
\theta}{m + \theta} \right)^2,
\enas
using the inequality $| ( 1 + x) e^{-x} -1 | \leq \frac{x^2}{2}$
for $x > 0$. The lemma now follows because $m+\theta >
m = 1+2\r$ and $1 + \theta>1$. 
 \ep

\medskip
Lemmas \ref{cont} and~\ref{phie} together yield the following result.

\begin{lemma} \label{phi-phie} 
For any $\r > 0$,
\beas
\| \phi_\r - \phi_e \|_{\cal G}
  &\leq& \frac{ \rho}{1 + 2\rho}.
\enas 
\end{lemma}

\proof
With Lemmas \ref{cont} and~\ref{phie}, it follows that
\beas
\| \phi_\r - \phi_e \|_{\cal G}  
    &=& \|\oper \phi_\r - \phi_e\|_{\cal G} \\
&\leq&  \| \oper \phi_\r - \oper \phi_e\|_{\cal G}
  + \| \oper \phi_e - \phi_e\|_{\cal G} \\
&\leq& \frac1m \| \phi_\r - \phi_e \|_{\cal G} 
     + \frac{2\rho^2}{(1 + 2 \rho)^2}.
 \enas
Note that indeed $\phi_\r - \phi_e
\in {\cal G}$.  Thus, since $m > 1$, it follows that  
\beas
 \| \phi_\r - \phi_e \|_{\cal G} &\leq&\frac{m}{m-1} \frac{2
   \rho^2}{(1 + 2 \rho)^2} = \frac{ \rho}{1 + 2 \rho},
\enas 
as required. 
\ep

As an immediate consequence, $\law(W_\r) \to {\rm NE}(1)$ as $\r\to0$.
Theorem~\ref{comp} reformulates this convergence as a pointwise
comparison theorem directly relevant to the distribution functions 
of $\D$ and~$T$. 

\begin{theorem} \label{comp}
Let $W,W'$ be independent~$\NE(1)$ random variables. Then,
for all $\theta>0$, we have
\beas
\left| \EE e^{- \theta W_\r {W_\r}'} -  \EE e^{- \theta W
W'}\right|&\leq& \frac{4 \rho}{1 + \rho}\, \theta^2.
 \enas 
\end{theorem}

\proof
We have
\beas
\lefteqn{\EE e^{- \theta W_\r {W_\r}'} 
  -  \EE e^{- \theta WW'}}\\
&=& \EE \{ \EE ( e^{- \theta W_\r W'_\r} 
   |W'_\r)\}  -  \EE\{ \EE ( e^{- \theta W W'} | W')\}\\
&=& \EE \phi_\r(\theta W'_\r) - \EE \phi_e (\theta W)\\
&=&  \EE \oper \phi_\r(\theta W'_\r) - 
   \EE \oper \phi_e (\theta W'_\r) + \EE \oper\phi_e(\theta W'_\r) 
     - \EE \phi_e (\theta W'_\r)\\ 
&&\qquad\mbox{} + \EE \phi_e(\theta W'_\r) - \EE \phi_e (\theta W). 
\enas 
Since 
\beas
\EE \phi_e (\theta W'_\r)&=& \EE e^{- \theta W W'_\r} 
  = \EE \phi_\r(\theta W),  
\enas
we obtain from the triangle inequality, \Ref{w-moments} and
Lemmas \ref{cont}, \ref{phie} and~\ref{phi-phie} that
\beas
\left| \EE e^{- \theta W_\r W'_\r} 
    -  \EE e^{- \theta WW'}\right|
&\leq& \frac{1}{m} \| \phi_\r - \phi_e \|_{\cal G}\th^2 \EE (W_\r^2) 
   + \frac{2 \rho^2}{(1 + 2 \rho)^2} \theta^2 \EE(W_\r^2)\\ 
&& \quad \mbox{} + \| \phi_\r - \phi_e \|_{\cal G}\theta^2 \EE (W_\r^2)\\
&\le& \frac{2\theta^2(1+\r)}{1+2\r}\left\{
  \Bl \frac1{1+2\r} + 1 \Br \frac{\rho}{1 + 2 \rho} 
   + \frac{ 2\rho^2}{(1 + 2 \rho)^2} \right\} \\
&\leq& \frac{4 \rho}{1 + 2\rho}\, \theta^2,
 \enas
as required.   
\ep

Noting that
$$
\EE e^{- \theta WW'} = \int_0^\infty {e^{-y} \over 1+\th y}\,dy,
$$
we obtain the following theorem. 

\begin{theorem} \label{corollary14}
As in Theorem \ref{corollary6}, let $\Delta$ be a random variable on~$\ZZ$ 
with distribution given by  
$$
\PP\left[\Delta > x \right]
= \EE \{  e^{-2 \phi_0^2(1+2\rho)^{x} W_\r {W_\r}'} \}.
$$
Let $T$ denote a random variable on~$\RR$ with
distribution given by  
\beas
\PP\left[T > z\right] = \int_0^\infty \frac{e^{-y}}{1 + 2ye^{2z}}\, dy.
\enas 
Then 
$$
\sup_{z \in \RR} |\PP[\r \Delta > z] - \PP[T  > z]|
  = O\Bl  \r^{1/3}(1 + \log(1/\r))  \Br.
$$
\end{theorem}

\proof    
 We use an argument similar to that used for Theorem~\ref{corollary6}. 
For~$a$ large, we can use the bound  
\begin{equation}\label{star-m}
\int_0^\infty \frac{e^{-y}dy}{1+ay} 
       \leq \int_0^1 {dy\over 1+ay} = a^{-1}\log(1+a), 
\en
from which, for $z>0$ and with $c(\r)$ defined by
$$
1 \ge c(\r) := (2\r)^{-1}\log(1+2\r) \ge 1 - \r,
$$
we have 
 \bea
\PP[T > zc(\r)] &\leq&   e^{-2zc(\r)}(1+zc(\r)) 
   \le (1+zc(\r))e^{-2z(1-\r)}.
\label{Tzcrho}
\ena  
Similarly, from Lemma~\ref{D-tail}, we have
\beas
\PP[\r \Delta > z] &\leq&  (1+2\r)^{-(z/\r)+2}(1 + zc(\r))
  \le (1+2\r)^2(1+zc(\r))e^{-2z(1-\r)}.
\enas

Complementing these upper tail bounds, from Theorem~\ref{comp}
and for $z \in \r\ZZ$, we have
\begin{equation}\label{body-diff-1}
\left| \PP[\r \Delta> z] - \int_0^\infty 
  {e^{-y} \over 1 + 2y\f_0^2(1+2\r)^{z/\r}}\,dy \right|
    \le {16\r \over 1+2\r}\,(1+2\r)^{2z/\r}
    \le {16\r \over 1+2\r}\,e^{4z}.
\end{equation}
Using the facts that $(1+2\r)^{z/\r} = e^{2zc(\r)}$ and that
$(1+2\r)^{-1} \le \f_0 \le 1$, and because, for $a,b>0$,
\begin{equation}\label{star-d}
\Blm \int_0^\infty {e^{-y} \over 1+ay}\,dy 
  - \int_0^\infty {e^{-y} \over 1+by}\,dy \Brm 
     \le {|b-a| \over \max\{1,a,b\}},
\end{equation}    
it also follows that
\bea\label{body-diff-2}
&&\Blm \int_0^\infty {e^{-y} \over 1 + 2y\f_0^2(1+2\r)^{z/\r}}\,dy
  - \PP[T_0 > zc(\r)] \Brm \nonumber\\
&&\qquad  \le 2|\f_0^2-1|(1+2\r)^{z/\r}
    \le {8\r\over 1+2\r}\,e^{2z};
\ena
and then, from \Ref{star-d} and~\Ref{star-m}, we have
\begin{equation}\label{body-diff-3}
|\PP[T > zc(\r)] - \PP[T > z]|
  \le \min\{4z(1-c(\r)),(2+z)e^{-z}\} = O(\r\log(1/\r)).
\en

Combining the bounds \Ref{body-diff-1}, \Ref{body-diff-2} and
\Ref{body-diff-3} for $e^{2z} \le \r^{-1/3}$ gives a supremum
of order $\r^{1/3}$ for $|\PP[\r \Delta > z] - \PP[T > z]|$;
note that~$z$ may actually be allowed to take any real value in this
range, since~$T$ has bounded density.  For any larger
values of~$z$, the upper tail bounds give a maximum
discrepancy of order $O\{\r^{1/3}(1 + \log(1/\r))\}$, as
required. Note that, in the main part of the distribution,
for~$z$ of order~$1$, the discrepancy is actually of order~$\r$.
\ep

\medskip
Numerically, instead of calculating the limiting distribution of 
$W_\r$, we would use the approximation
\beas
&&\left| 
  \EE \left\{e^{- L^{-1}(N_{r'}s_r+M_ru_{r'})} \right\} - 
  \EE \left\{ e^{-2 \phi_0^2(1+2\rho)^{r+r'}  
                 W(\tau_r) W'(\tau_{r'})}\right\} \right| \\ 
&&\qquad\leq 6\phi_0^{3/2}(1+\r)(1+2\rho)^{r+r'- \frac12(r \wedge r')} 
    (L\r)^{-1/4} ,
\enas
from \Ref{star-1} and~\Ref{term-1},
where the distributions of $W(\tau_r)$ and $W'(\tau_{r'})$ can be calculated
iteratively, using the generating function from Lemma~\ref{2.01}. 
As $D$ is centred around $2 n_0 = 2 \lfloor
\frac{N}{2}\rfloor$, and as $r$ is of order at most $\frac{\log(L \rho)
}{\log(1 + 2 \rho)}$,  only order  $\frac{\log(L \rho) }{\log(1 + 2 \rho)}$
iterations would be needed.

\section{The discrete circle model: description}  
\setcounter{equation}{0} 

Now suppose, as in the discrete circle model of 
Newman {\it et al.\/}~\cite{NewmanWatts2}, that 
the circle~$C$ becomes a ring lattice with~$\L = Lk$ vertices, where 
 each vertex is  connected to all its neighbours within
distance~$k$ by an undirected edge. In the notation of~\cite{NewmanWatts2}, 
a number of shortcuts are added
between randomly chosen pairs of sites, with probability $\phi$ per
connection in the lattice, of which there are $\Lambda k$; thus, on
average, there are $\L k\phi$ shortcuts in the graph. In contrast to 
the previous setting, it is natural in the discrete model to use graph 
distance, which implies that all edges, {\it including shortcuts\/}, 
have length~1. 
This turns out to make a significant difference to
the results when shortcuts are very plentiful.

For ease of comparison with the previous model, which collapsed the 
$k$-neighbourhoods, we 
adopt a different notation. The model can be formulated as the union
of a Bernoulli random graph $G_{\Lambda,\frac{\sigma}{\L}}$ and 
the underlying ring
lattice on~$\Lambda$ vertices. Here  we write $\sigma = \frac{\rho}{k}$, 
so that the expected number of edges in $G_{\Lambda, \frac{\sigma}{\L}}$
is close to the value $L\r/2$ in the previous model; comparing
the expected number of shortcuts with that given in~\cite{NewmanWatts2},
we also have
$$
\L k\f = \half \Lambda (\L-2k-1)\frac{\sigma}{\L} = \frac\r{2k}(\L-2k-1)
\approx \half L\r,
$$
relating our parameter~$\s$ to those of~\cite{NewmanWatts2}. In particular,
we have
\bea \label{rhoscale}
&&\s = \frac{2 \L k \phi}{\L-2k-1} \approx 2 k \phi.  
\ena 

The model can also be realized by a dynamic construction.
Choosing a point $P_0 \in \{1, \ldots, \L\}$ at random, set $R(0)=\{P_0\}$. 
Then, at the first step (distance~1), the `island' consisting of~$P_0$
is increased by $k$ points at each end, and, in addition, 
$M_1^{(1)}\sim\Bi(\L-2k-1,\frac{\sigma}{\L})$ shortcuts connect to 
centres of new islands.  
At each subsequent step,  starting from the set~$R(n)$ of vertices
within distance~$n$ of~$P_0$, each island is increased by the addition
of~$k$ points at either end, but with overlapping islands merged, to
form a set~$R'(n+1)$; this is then increased to~$R(n+1)$ by choosing
a Bernoulli--$\frac{\sigma}{\L}$ thinning of the edges joining 
$R(n)\setminus R(n-1)$
to $C\setminus R'(n+1)$ as shortcuts.

The branching analogue of this process, which agrees with the current
process until its first self--overlap occurs, has individuals, here
representing the islands, of two types: newly formed type~1 islands, 
consisting of just one vertex, and existing type~2 islands. A type~1 
island at time~$n$ becomes a type~2 island at time $n+1$, and,
in addition, has a $\Bi(\L,\frac{\sigma}{\L})$--distributed number of type~1 
islands as `offspring'. A
type~2 island at time~$n$ stays a type~2 island at time $n+1$, and 
has a $\Bi(2k\L,\frac{\sigma}{\L})$--distributed number of type~1 
islands as offspring.
Each new island starts at an independent and randomly chosen
point of the circle,
and at each subsequent step acquires~$k$ more vertices at either end.
Writing 
$$
\hm(n):= ({\hat M}^{(1)}(n), {\hat M}^{(2)}(n))^T, \quad{n \geq 0},
$$
for the numbers of islands of the two types at time~$n$, 
where the superscript $T$ denotes the transpose, their
development over time is given by the branching recursion 
\bea 
\hm^{(1)}(n) &\sim& \Bi \left((\hm^{(1)}(n-1) + 2 k \hm^{(2)}(n-1))\L, 
    \frac{\sigma}{\L}\right),\non\\  
\hm^{(2)}(n)&=& \hm^{(1)}(n-1) + \hm^{(2)}(n-1):\non\\
\hm^{(1)}(0)&=&1, \hh \hm^{(2)}(0)=0.\label{IC}
\ena 
The total number of intervals at time $n$
is denoted by
\bea\label{mplusn}
\hm^+(n) =
\hm^{(1)}(n) + \hm^{(2)}(n),
\ena
and the total number of
vertices in these intervals by
\bea \label{sn}
\hs(n) = \hm^+(n) + 2k \sum_{j=0}^{n-1} \hm^+(j) \ge \hm^+(n). 
 \ena 
 
As before, we use the branching process as the basic tool in our
argument. It is now a two type
Galton--Watson process with mean matrix  
\beas
A = \left( 
\begin{array}{cc} 
\s & 2 k \s \\
1 & 1
\end{array}
\right).
\enas
The characteristic equation
\bea \label{ev}
(t - 1)(t - \s) = 2k\s
\ena
of $A$ yields the eigenvalues  
\beas \lambda \,=\, \lambda_1 &= &\half\{ \s + 1 +
   \sqrt{(\s + 1)^2 + 4 \s (2k-1)}\} > \s + 1;\\
-\l \,<\,\lambda_2 &= &\half\{ \s + 1 -
   \sqrt{(\s + 1)^2 + 4 \s (2k-1)}\} <0: 
\enas
also, from~\Ref{ev},
\bea \label{1+2}
\lambda + \lambda_2 = \s +1 \quad \mbox{and}\quad \l\l_2
  = -\s(2k-1).
\ena 

From the equation $f A = \lambda f$, we find that the left
eigenvectors~$f^{(i)}$, $i=1,2$, satisfy
\bea \label{f1f2}
f_2^{(i)} = (\lambda_i - \s) f_1^{(i)}.  
\ena 
We standardize the positive left eigenvector~$f^{(1)}$ of~$A$,
 associated with the eigenvalue~$\lambda$,
so that 
\bea
 f_1^{(1)} &=& (\lambda - \s)^{-\frac12}, \hh f_2^{(1)} \,=\,
(\lambda - \s)^\frac12 ; \label{f11-def}
\ena
for $f^{(2)}$, we choose
\beas
 f_1^{(2)}= (\s - \lambda_2)^{-\frac12}, \hh f_2^{(2)} =
- (\s - \lambda_2)^\frac12 .
\enas 

\medskip
Then, for $i=1, 2$,  we have 
\beas
\EE ( (f^{(i)})^T \hm_{n+1} | {\cal F}(n))&=&   (f^{(i)})^T A \hm(n) =
\lambda_i  (f^{(i)})^T \hm(n),
 \enas 
where
${\cal F}(n)$ denotes the $\s$-algebra 
$\sigma(\hm(0), \ldots, \hm(n))$. Thus, from (\ref{f1f2}), 
\bea \label{wni}
W^{(i)}(n) &:=& \lambda_i^{-n}  (f^{(i)})^T \hm(n)\\
&=&\lambda_i^{-n} f_1^{(i)} ( \hm_{n}^{(1)} +(\lambda_i - \s) \hm_{n}^{(2)}
)  \nonumber 
\ena 
is a (non-zero mean) martingale, for $i=1,2$; we let 
\bea \label{wrho}
W_{k,\s} := \lim_{n \rightarrow \infty} W^{(1)}(n) {\mbox { a.s.} } 
= \lim_{n \rightarrow \infty}\lambda_1^{-n}  (f^{(1)})^T \hm(n)   
{\mbox { a.s.} }
\ena 
be the almost sure limit of the martingale $W^{(1)}(n)$.

Our main conclusions can be summarized as follows: the detailed results
and their proofs are given in Theorems \ref{corollary6d}
 and~\ref{drhosmall}.
Let $\Delta_d$ denote a random variable on the integers with 
distribution given by
\bea
\PP\left[\Delta_d > x \right]= 
\EE\exp  \Blb - \messls 
     \f_d^2 \l^x W_{k,\s} W'_{k,\s} \Brb , \label{Delta-d-dist}
\ena
for any $x \in \ZZ$,
and set $D^* = \D_d + 2n_d  $. Here, $n_d$ and $\phi_d$ are such that
$\l^{n_d} = \f_d(\L\s)^{1/2}$ and $\l^{-1} < \f_d \le 1$. 
Let~$D$ denote the graph distance between a randomly chosen pair
of vertices on the ring lattice~$C$.

\begin{theorem}\label{summary-2}
If $\L\s \to \infty$ and $\r = k\s$ remains bounded, then it
follows that
$\dtv({\cal L}(D), {\cal L}(D^*)) \to 0$.
If $\r \rightarrow 0$, then $\r\Delta_d \to_{{\cal D}}T$, where~$T$ is as  
in Theorem~\ref{summary}.
\end{theorem}

\nin Note that the expectation in \Ref{Delta-d-dist} is taken under
the initial condition~\Ref{IC}; we shall later need also to consider
the distribution of~$W_{k,\s}$ under other initial conditions.

\section{The discrete circle model: proofs}  
\setcounter{equation}{0}

We begin the detailed discussion with some moment formulae.

\begin{lemma}\label{moments}
For the means,
\bea 
\EE \hm^{(1)}(n) &=& \frac{1}{\lambda - \lambda_2} ( \lambda^n ({\s -
    \lambda_2}) + \lambda_2^n (\lambda- \s ) ); \nonumber\\
\EE \hm^{(2)}(n) &=&  \frac{1}{\lambda - \lambda_2} ( \lambda^n
    - \lambda_2^n); \nonumber\\
\EE \hm^+(n) &=& \frac{1}{\lambda - \lambda_2} (\lambda^{n+1} -
    \lambda_2^{n+1} ) \le 2\l^n,  \label{ehm}
\ena
and
\bea   
\EE(\hm_{n}^{(1)} + 2 k \hm_{n}^{(2)} ) &=&
 \frac{1}{(\lambda - \lambda_2)}  \{ (1 - \lambda_2/\s) \lambda^{n+1} +
  (\l_2/\s)(\lambda - \s) \lambda_2^{n} \} \nonumber \\ 
&\leq&  c \lambda^n, \mbox{  for  } c = 4k-1. \label{ebnd} 
\ena  
For the variances, for $j \le n$, 
\bea  
\var (W^{(1)}(j) -  W^{(1)}(n))\leq  
 \k^2 (f_1^{(1)})^2 \lambda^{-j},\qquad \k^2 := {c\s\over\l(\l-1)};
 \qquad&&
 \label{inc1} \\
\var (W^{(2)}(j) -  W^{(2)}(n))\  
  \phantom{\leq\k^2 (f_1^{(1)})^2 \lambda^{-j},\qquad 
    \k^2 := {c\s\over\l(\l-1)};}\qquad&&
      \label{inc2} \\
\qquad \le c \s \Bl{f_1^{(2)}\over\l_2}\Br^2 
   \min\Blb{\l_2^2\over| \lambda - \lambda_2^2|},(n-j)\Brb
   \left(\frac{\lambda}{\lambda_2^2}\right)^j 
   \max \Blb 1, \frac{\lambda}{\lambda_2^2}\Brb^{n-j}&&
 \non
\ena 
and, for $\hm^+(n)$, 
\bea 
\label{varhm} 
{\var \hm^+(n)}
&\leq &  4\k^2 \lambda^{2n};\\
\label{mnmn} 
\EE \hm^+(n)(\hm^+(n) - 1) &\leq & 
4(1+\k^2)
\lambda^{2n}. 
 \ena 
\end{lemma}

\nin Note that, from~\Ref{ev}, we have 
\bea \label{kappabound} 
&&0 \le \k^2 =  \frac{c}{2k+\l-1} \le \frac{4k-1}{2k} \le 2. 
\ena

\proof
 First, observe that 
\begin{equation}\label{ew}
\EE W^{(i)}(n) = W_0^{(i)} =   (f^{(i)})^T \hm_0^+ 
  = (f^{(i)})^T (1,0)^T = f_1^{(i)}
\end{equation} 
for all~$n$, by the martingale property. 
From (\ref{wni}) and (\ref{f1f2}), we have  
\beas
(f_1^{(1)})^{-1}\lambda^n W^{(1)}(n)&=& \hm^{(1)}(n) + (\lambda - \s)
  \hm^{(2)}(n);\\
(f_1^{(2)})^{-1}\lambda_2^n W^{(2)}(n)&=& 
   \hm^{(1)}(n) + (\lambda_2 - \s) 
\hm^{(2)}(n),
\enas 
and thus
\bea \label{meq}
\hm^{(1)}(n) &=& \l^n W^{(1)}(n)\frac{\s - \lambda_2}{(\l -
   \lambda_2) f_1^{(1)} } + \lambda_2^n  W^{(2)}(n) \frac{\l-
    \s}{(\l - \lambda_2) f_1^{(2)} }; \\
\hm^{(2)}(n) &=& \l^n W^{(1)}(n)\frac{1}{(\l -
   \lambda_2) f_1^{(1)} } - \lambda_2^n  W^{(2)}(n) \frac{1}{(\l -
     \lambda_2) f_1^{(2)} } \nonumber. 
 \ena
From (\ref{meq}) and (\ref{ew}) we obtain
\beas
\EE \hm^{(1)}(n) &=& \l^n \frac{\s - \lambda_2}{(\l -
\lambda_2) } + \lambda_2^n \frac{\l-\s }{(\l - \lambda_2) }; \\
\EE \hm^{(2)}(n) &=& \l^n \frac{1}{(\l -
\lambda_2) } - \lambda_2^n \frac{1}{(\l -
\lambda_2)},
 \enas 
giving (\ref{ehm}); for the last part use  $\s + 1 - \l=
\lambda_2$ and  $\s+1-\lambda_2 = \l$, from~\Ref{1+2}. 
Then (\ref{ebnd}) follows immediately,
using  (\ref{ev}) and (\ref{1+2}).

Now define 
\bea \label{xn}
X(n) := \hm^{(1)}(n) - \s (\hm^{(1)}(n-1) + 2 k\hm^{(2)}(n-1)),
\quad n\ge1,
\ena 
noting that it has a centred binomial distribution conditional on
${\cal F}(n-1)$; representing quantities in terms of these
martingale differences greatly simplifies the subsequent
calculations.  For instance,   
\bea \label{nice} 
&&W^{(i)}(n+1) - W^{(i)}(n) \nonumber \\
&&\quad= \lambda_i^{-n-1} f_1^{(i)}
   \{\hm^{(1)}(n+1) + (\l_i-\s)\hm^{(2)}(n+1) \nonumber \\
&&\qquad\qquad  \mbox{}  - \l_i\hm^{(1)}(n)
    -\l_i(\l_i-\s)\hm^{(2)}(n)\}\nonumber \\ 
&&
\quad = \lambda_i^{-n-1} f_1^{(i)}\{\hm^{(1)}(n+1)
    -\s\hm^{(1)}(n) - 2k\s\hm^{(2)}(n)\}\nonumber  \\ 
&&
\quad = \lambda_i^{-n-1} f_1^{(i)} X(n+1),
\ena 
where we have used $(\l_i-1)(\l_i-\s) = 2k\s$, from~\Ref{ev}, 
and the branching recursion. 

Since
$$
\EE \{ X^2(n+1)  \giv {\cal F}(n)\} = \frac\s\L \left( 1 -
  \frac\s\L  \right) (\hm^{(1)}(n) + 2 k \hm^{(2)}(n))\L,
$$ 
we have
$$
\EE X^2(n+1) \le c\s\l^{n},
$$
from~\Ref{ebnd}.  Thus, immediately, 
\begin{equation}\label{var12} 
\EE \{ (W^{(i)}(n+1) - W^{(i)}(n) )^2 \}
\leq c \s (f_1^{(i)})^2  \lambda_i^{-2n-2}  \l^n.  
\end{equation}
Hence, for $i=1, 2$ and for any $0\le j < n$, 
\beas
&&\var (W^{(i)}(j) -  W^{(i)}(n)) \\ 
&&\qquad =  \sum_{k=j}^{n-1}\EE (W^{(i)}(k) - W^{(i)}(k+1))^2 \\
&&\qquad\leq c \s (f_1^{(i)})^2  
   \sum_{k=j}^{n-1} \lambda^k\lambda_i^{-2(k+1)}\\
&&\qquad\leq c \s (f_1^{(i)})^2 \l_i^{-2}
   \left( \frac{\lambda}{\lambda_i^2}\right)^j 
   \min\left\{\frac{\l_i^2}{| \lambda - \lambda_i^2| },(n-j)\right\} 
   \max \left( 1,\left(\frac{\lambda}{\lambda_i^2}\right)^{n-j} \right),
\enas
and the formulae \Ref{inc1} and~\Ref{inc2} follow.

Moreover, from~\Ref{meq},
\bea\label{one-star}
\hm^+(n) =\frac{1}{\lambda - \lambda_2}\left( 
\frac{\lambda^{n+1}}{f_1^{(1)}} W^{(1)}(n) -
\frac{\lambda_2^{n+1}}{f_1^{(2)}} W^{(2)}(n)\right) , 
\ena
and hence   
 \beas
(\lambda - \lambda_2)^2 \var \hm^+(n) 
&\qquad= &\sjn(\l^{n+1-j} - \l_2^{n+1-j})^2\var X(j) \\
&\qquad \le&  4c\s\sjn \l^{2n-(j-1)} \le 4c\s\l^{2n+1}/(\l-1).   
\enas   
From this,  using the inequality 
\beas
\EE \hm^+(n)(\hm^+(n) - 1) &\leq & \var(\hm^+(n)) + (\EE \hm^+(n))^2, 
\enas 
(\ref{mnmn}) is easily obtained. \ep

As in the previous section, we run two branching processes $\hm$
and $\hm'=:\hn$ independently, and investigate the time at which
the first intersection occurs, irrespective of the types of the
intervals.  We write $\hs'(n) =: \hu(n)$, and use notation of the
form~$\hm_r$ to denote $\hm(n_d+r)$, for an appropriate~$n_d$
which we shall define later;
we also use $\t_r := \{2k(n_d+r) + 1\}$ to denote the length of
the longest interval in the branching process at time $n_d+r$.
Then, with~$\whv_{r,r'}$ defined as before to be the number of
pairs of intervals of $\hm$ and~$\hn$ intersecting, when~$\hm$
has been run for time $n_d+r$ and~$\hn$ for time $n_d+r'$, the
analogue of Proposition~\ref{cor3} shows that
\bea
&&|\PP[\whv_{r,r'} = 0] - \PP[D > 2n_d + r + r']|\non\\
&&\qquad \le 32 \L^{-2}\t^2_{(r\vee r')}\EE\{\half
  \hm^+_r\hn^+_{r'}(\hm^+_r + \hn^+_{r'} - 2)\},\label{C-3}
\ena
and that of Corollary~\ref{cor2} gives
\bea
&&|\PP[\whv_{r,r'} = 0 \giv \hm^+_r=M,\hn^+_{r'}=N,\hs_r=t,\hu_{r'}=u]\non\\
&&\hskip1in \mbox{} - \exp\{-\L^{-1}(Nt+Mu-MN)\}| \non \\
&&\qquad \le 8\L^{-1}(M+N)\t_{(r\vee r')}.\label{C-2}
\ena    
The estimates \Ref{C-3} and~\Ref{C-2} can be made more explicit
with the help of the bounds
\begin{equation}\label{C-1}
\EE \hm^+_r \le 2\l^{n_d+r};\qquad 
  \EE\hm^+_r(\hm^+_r-1) \le 4(1+\k^2)\l^{2(n_d+r)},
\end{equation}
which follow from from Lemma~\ref{moments}; together, they
give the following result, in which~$D$ denotes the shortest
distance between~$P_0$ and a randomly chosen vertex~$P'$ of~$C$. 

\begin{lemma}\label{cor4d}
With the above notation and definitions,  we have 
\beas
&&|\PP[D > 2n_d + r + r'] - \PP[\whv_{r,r'} = 0]|\\
&&\qquad \le 256 \L^{-2}\t^2_{(r\vee r')}(1+\k^2)
    \l^{3n_d+r+r'+(r\vee r')},
\enas
and
\beas
&&|\PP[\whv_{r,r'} = 0 \giv \hm^+_r=M,\hn^+_{r'}=N,\hs_r=t,\hu_{r'}=u]\\
&&\hskip1in \mbox{}
  - \exp\{-\L^{-1}(Nt+Mu-MN)\}| \\
&&\qquad \le 32\L^{-1}\t_{(r\vee r')}\l^{n_d+(r\vee r')}.
\enas 
\end{lemma}

\nin  We now need to examine 
$\EE\exp\{-\L^{-1}(\hn^+_{r'}\hs_r+\hm^+_r\hu_{r'}-\hm^+_r\hn^+_{r'})\}$
more closely.

To start with, from  (\ref{ehm}) in Lemma \ref{moments}, we have
\bea  
\EE {\hat s}(n) &=& \frac{1}{\l - \lambda_2}\left\{
  (\l^{n+1}-\l_2^{n+1}) + 2k\sum_{j=0}^{n-1}(\l^{j+1}-\l_2^{j+1})\right\}
  \non \\ 
&=& \frac{1}{\l - \lambda_2} \left\{ 
  \l^{n+1}  \left( 1 + \frac{2k}{\l-1} \right) - 
  \lambda_2^{n+1}  \left( 1 + \frac{2k}{\lambda_2-1} \right)\right.\non \\
&&\left.\qquad\qquad\qquad \mbox{}
     - 2k\left(\frac{\l}{\l-1} - \frac{\lambda_2}{\lambda_2-1}  
   \right) \right\}\nonumber  \\
&=& \frac{1}{\s(\l - \lambda_2) } \left\{ \l^{n+2}   - 
 \lambda_2^{n+2}  - (\l-\l_2) \right\}, \label{esn}
\ena 
where we have used (\ref{ev}) and~\Ref{1+2} to simplify, and
this expression is rather close to $(\l/\s)\EE \hm^+(n)$ as given
in~(\ref{ehm}).  This reflects the fact that both $\hs(n)$
and~$(\l/\s)\hm^+(n)$ are rather close to
$$
{\l^{n+2}\over\s(\l-\l_2)}\,{W\ui(n-1)\over f_1\ui}.
$$

\begin{lemma}\label{s_and_Mplus}
We have the following approximations:
\beas
\hs(n) &=&
{\l^{2+n}\over\s(\l-\l_2)}\Bl {W\ui(n-1)\over f_1\ui} + \tu_1(n)\Br;\\
{\l\over\s}\hm^+(n) &=&
{\l^{2+n}\over\s(\l-\l_2)}\Bl {W\ui(n-1)\over f_1\ui} + \tu_2(n)\Br,
\enas
where
\bea
\EE|\tu_1(n)| &\le& \{3 + 2\k\sqrt{n+\s^2}\}
   \max \Blb \l^{-1/2},{|\l_2|\over\l}\Brb^{n}; \label{A1-d}\\
\EE|\tu_2(n)| &\le& \{1 + 2\k\sqrt{n+\l-1}\}
   \max \Blb \l^{-1/2},{|\l_2|\over\l}\Brb^{n}. \label{A2-d}
\ena 
\end{lemma}    
   
\proof
We first
express $\hs(n)$ and~$(\l/\s)\hm^+(n)$ in terms of the martingale
differences $\{X(l),\,l\ge1\}$. 
From \Ref{one-star} and~(\ref{nice}), we have
\beas
\hm^+(n)& =& \frac{1}{\l - \l_2} \left\{ \frac{\l^{n+1}}{ f_1\ui}W\ui(n) 
   -  \frac{\l_2^{n+1}}{f_1^{(2)}} W^{(2)}(n)\right\} \\
&=& \frac{\l^{n+1}}{(\l - \l_2) f_1\ui}W\ui(n-1) + X(n) 
- \frac{\l_2^{n+1}}{\l - \l_2} 
   \left(1 + \sum_{l=1}^{n-1} \l_2^{-l}X(l) \right). 
\enas 
Similarly, from \Ref{one-star} and~(\ref{nice}), 
\beas
\lefteqn{\sum_{j=0}^{n-1} \hm^+(j)}\\
 &=& \frac1{\l-\l_2}\sum_{j=0}^{n-1}\Blb \frac{\l^{j+1}}{f_1\ui}W\ui(j)
 - \frac{\l_2^{j+1}}{f_1\ut}W\ut(j)\Brb\\
&=&  \frac1{\l-\l_2}\sum_{j=0}^{n-1}\Blb \frac{\l^{j+1}}{f_1\ui}W\ui(n-1)
 - \l^{j+1}\sum_{l=j+1}^{n-1}\l^{-l}X(l) \right.\\
&&\qquad\qquad\quad \left. \mbox{}
    - \l_2^{j+1}\Bl 1 + \sum_{l=1}^j\l_2^{-l}X(l)\Br\Brb\\
&=&  \frac1{\l-\l_2}\Blb \frac{ W\ui(n-1)}{f_1\ui}\,\frac{\l^{n+1}}{\l-1}
  - \frac\l{\l-1}\Bl 1 + \sum_{j=1}^{n-1}\l^{-j}X(j) \Br
     \right.\\
&&\qquad\qquad\quad \left. \mbox{} 
    - \sum_{l=1}^{n-1} \l^{-l}X(l) \sum_{j=0}^{l-1}\l^{j+1}
      - \frac{\l_2^{n+1}-\l_2}{\l_2-1}
        -\sum_{l=1}^{n-1}\l_2^{-l}X(l)\sum_{j=l+1}^{n}\l_2^{j}\Brb \\
&=&  \frac1{\l-\l_2}\Blb \frac{ W\ui(n-1)}{f_1\ui}\,\frac{\l^{n+1}}{\l-1}
 - \Bl \frac\l{\l-1} - \frac{\l_2}{\l_2-1} \Br 
     \Bl 1 + \sum_{l=1}^{n-1}X(l) \Br \right.\\
&&\qquad\qquad\quad \left. \mbox{} 
    - \frac{\l_2^{n+1}}{\l_2-1}
         \Bl 1 + \sum_{l=1}^{n-1}\l_2^{-l}X(l) \Br\Brb.
\enas        
Substituting these into~\Ref{sn}, and because $1 + 2k/(\l_i-1) = \l_i/\s$,
$i=1,2$, 
from~\Ref{ev} and~\Ref{1+2}, we obtain
\beas
\hs(n) &=& {\l^{2+n}\over\s(\l-\l_2)}\,{W\ui(n-1)\over f_1\ui} + X(n)\\
&&\qquad \mbox{} - {1\over\s}\Blb \sum_{l=1}^{n-1} X(l)
    \Bl 1 + {\l_2^{n+2-l}\over \l-\l_2}\Br 
       + 1 + {\l_2^{n+2} \over  \l-\l_2} \Brb \\
 &=&  {\l^{2+n}\over\s(\l-\l_2)}\Bl {W\ui(n-1)\over f_1\ui} + \tu_1(n)\Br,
\enas
where
$$
|\EE\tu_1(n)| = \l^{-n-2}|\l-\l_2+\l_2^{n+2}| 
  \le 3 \max \Blb \l^{-1},{|\l_2|\over\l}\Brb^{n+1}
$$
and
\beas
\var\tu_1(n) &\le& \l^{-2n-4}c\s(\l-\l_2)^2
  \Blb \s^2\l^{n-1} + \sum_{l=1}^{n-1} \l^{l-1} 
    \Bl 1 + {\l_2^{n+2-l}\over \l-\l_2}\Br^2 \Brb  \\
&\le& 4\k^2  (\s^2 + n) 
    \max \Blb \l^{-1},{\l_2^2\over\l^2}\Brb^{n+1},
\enas
giving the first approximation.
By similar arguments, for $(\l/\s)\hm^+(n)$  we obtain
\beas
{\l\over\s}\hm^+(n) &=& {\l^{2+n}\over\s(\l-\l_2)}\,{W\ui(n-1)\over f_1\ui}\\ 
&&\qquad \mbox{} +{\l\over\s}\Blb X(n) 
   - {1\over\l-\l_2} \sum_{l=1}^{n-1} X(l)\l_2^{n+1-l} 
       - {\l_2^{n+1} \over  \l-\l_2}\Brb  \\
 &=&  {\l^{2+n}\over\s(\l-\l_2)}\Bl {W\ui(n-1)\over f_1\ui} + \tu_2(n)\Br,
\enas
where
$$
|\EE\tu_2(n)| = (|\l_2|/\l)^{n+1}  
$$
and   
\beas
\var\tu_2(n) &\le& \l^{-2n-2}c\s(\l-\l_2)^2
  \Blb \l^{n-1} + \sum_{l=1}^{n-1} {\l^{l-1} 
    \l_2^{2(n+1-l)}\over (\l-\l_2)^2} \Brb  \\
&\le& 4\k^2  (\l - 1 + n) 
    \max \Blb \l^{-1},{\l_2^2\over\l^2}\Brb^{n},
\enas
giving the second approximation. \ep

We now use these approximations as in the previous section, 
starting by observing that 
\beas
\lefteqn{\e(n,n') := \Blm 
 \EE\exp\{-\L^{-1}(\hn^+(n')\hs(n)+\hm^+(n)\hu(n')-\hm^+(n)\hn^+(n'))\}
 \phantom{\Bl{\l^{n+n'}\over L\r}\Br HHHHHHHHHHHHH}
   \right.}\\
&& \left.\mbox{} - 
  \EE \exp \Blb -\messls
   \Bl{\l^{n+n'}\over \L\s}\Br
    W\ui(n-1)W\uid(n'-1)\Brb\Brm \phantom{HHHHHH}\\
&\le& {\l^3\over\s(\l-\l_2)^2}\,{\l^{n+n'}\over \L}
  \Blb \EE|\tu'_1(n')|\EE\{\l^{-n}\hm^+(n)\} + \EE|\tu_2(n)|\right. \\    
&&\hskip1.5in \left. \mbox{} +  
   \EE|\tu_1(n)|\EE\{\l^{-n'}\hn^+(n')\} + \EE|\tu'_2(n')|\Brb \\
&&\quad \mbox{} + \Bl {\l\over\l-\l_2} \Br^2\, {\l^{n+n'}\over \L}
  \Blb  \EE|\tu_2(n)|\EE\{\l^{-n'}\hn^+(n')\} + \EE|\tu'_2(n')|\Brb ,
\enas
since $\EE W\ui(m) = f_1\ui = (\l-\s)^{-1/2}$ for all~$m$.  
Since also, from~\Ref{ehm}, 
$\l^{-m}\EE\hm^+(m) \le 2$, it follows from Lemma~\ref{s_and_Mplus}
and because $\l\ge 1+\s > 1$ that 
\bea
\e(n,n') &\le&  \{17 + 18\k\sqrt{(n\vee n')+(\s^2\vee(\l-1))}\}
   {\l^3\over(\l-\l_2)^2}\,{\l^{n+n'}\over \L\s} \non\\
 &&\qquad \mbox{}\times
   \max \Blb \l^{-1/2},{|\l_2|\over\l}\Brb^{(n\wedge n')}.
     \label{first-split}
\ena
Then similarly, since $(W\ui(j)-W_{k,\s})/f_1\ui$ has mean zero
and, letting $n\to\infty$ 
in~\Ref{inc1}, variance at most $\k^2\l^{-j}$, it follows that
\bea
\lefteqn{\Blm \EE \exp \Blb -\messls
     \Bl{\l^{n+n'}\over \L\s}\Br
       W\ui(n-1)W\uid(n'-1)\Brb \right.}\non\\
&&\qquad\qquad \left. \mbox{} 
   - \EE\exp  \Blb -\messls
     \Bl{\l^{n+n'}\over \L\s}\Br
       W_{k,\s} W'_{k,\s}\Brb \Brm \non\\ 
&\le& 2\messls
     \Bl{\l^{n+n'}\over \L\s}\Br\,
       \k\l^{-\{(n\wedge n')-1\}/2}.\label{second-split}
\ena

So choose $n_d$ so that
$\l^{n_d} = \f_d(\L\s)^{1/2}$ with $\l^{-1} < \f_d \le 1$, and let
$n=n_d+r$, $n'=n_d+r'$; then define the quantities
\beas
\h'_1(r,r') &:=& 256\f_d^3\{\s(2k(n_d+(r\vee r'))+1)\}^2(1+\k^2)
   \l^{r+r'+(r\vee r')};\\
\h'_2(r,r') &:=& 32\f_d\{\s(2k(n_d+(r\vee r'))+1)\}
   \l^{(r\vee r')};\\ 
\h'_3(r,r') &:=& 4\f_d^{3/2}\l^{1/2}(\l-\s)^2\k
   \l^{r+r'-(r\wedge r')/2},
\enas
and
$$
\h'_4(r,r') := 18\f_d^{2-\g}\l^{1-\g}
  \{1 + \k(n_d+(r\vee r')+(\s^2\vee(\l-1)))^{1/2}\}
     \l^{r+r' - \g(r\wedge r')},
$$
where
\bea \label{gamma}
\g &:=& \g(k,\s) := \min\{\half,(\log(\l/|\l_2|)/\log\l\}.
\label{gamma-def}
\ena
Note that, for fixed $k\s = \r$, simple differentiation shows that
$\l_1$ is an increasing function of~$\s$ and $|\l_2|$ a decreasing
function, so that $\l_1 (\s) \ge \l_1 (0)$, $|\l_2(\s)| \le |\l_2(0)|$,
and hence
$$
\frac{\log(\l/|\l_2|)}{\log\l} = 1 - \frac{\log|\l_2|}{\log\l} \ge 1 
   - \frac{\log(\sqrt{1+8\r}-1)}{\log(\sqrt{1+8\r}+1)} \ge \frac12
$$
in $\r\le1$. 
Thus, for $\r \le 1$, we have $\gamma=\frac12$.

Then, from Lemma~\ref{cor4d} and \Ref{first-split}
and~\Ref{second-split}, we have the following analogue
of Theorem~\ref{D4}.

\begin{theorem}\label{D4-a}
With the above assumptions and definitions, for $x\in\ZZ$
and $r=r(x)=\lfloor x/2\rfloor$, $r'=r'(x)=x-r(x) \le (x+1)/2$, we have
\beas
&&\Blm \PP[D > 2n_d+x] - \EE\exp  \Blb - \messls
     \f_d^2\l^x W_{k,\s} W'_{k,\s}   \Brb \Brm \\
&&\qquad \le (\h'_1(r,r') + \h'_2(r,r'))(\L\s)^{-1/2}
   + \h'_3(r,r')(\L\s)^{-1/4} + \h'_4(r,r')(\L\s)^{-\g/2}.
\enas 
In particular, if $\rho \le 1$, 
\beas
&&\Blm \PP[D > 2n_d+x] - \EE\exp  \Blb - \messls
     \f_d^2\l^x W_{k,\s} W'_{k,\s}\Brb \Brm \\
&&\qquad \le (\h'_1(r,r') + \h'_2(r,r'))(\L\s)^{-1/2}
   + (\h'_3(r,r') + \h'_4(r,r')) (\L\s)^{-1/4}.
\enas 
\end{theorem}

\nin Note that the expectation in Theorem~\ref{D4-a} is taken
conditional on the initial condition $\hm_0 = {\bf e}\ui$.

The theorem can be translated into a uniform bound, similar to that of
Theorem~\ref{corollary6}.  To do so, we need to be able to
control $\EE\{e^{-\ps W_{k,\s}W'_{k,\s}}\}$ for large~$\ps$.
The following analogue
of Lemma \ref{D-tail} makes this possible. To state it, we
first need some notation. 

For $W_{k,\s}$ as in
(\ref{wrho}), let
 $\phi_{k,\s} := (\phi_1, \phi_2)$ denote
 the Laplace transforms 
\bea \label {phi}
\phi_1(\theta) &=& \EE\{ e^{- \theta (f_1^{(1)})^{-1} W_{k,\s}}\giv
 \hm_0 = {\bf e}\ui\} ;\\
\phi_2(\theta) &=& \EE\{ e^{- \theta  (f_1^{(1)})^{-1} W_{k,\s}}\giv
\hm_0 = {\bf e}\ut\} 
\nonumber 
\ena
of ${\cal L}(  (f_1^{(1)})^{-1} W_{k,\s})$, where ${\bf e}^{(i)}$ is the
$i$'th unit vector.  Although we now
need to distinguish other initial conditions for the branching process,
 {\it unconditional\/} expectations
will always in what follows presuppose the initial condition 
$\hm_0 = {\bf e}\ui$, as
before. Then, as in Harris~\cite{Harris},  p.45,
$\phi_{k,\s}$ satisfies the Poincar\'e equation
\bea\label{Poincare}
 \phi_i( \lambda \theta) &=& g^i(\phi_1(\theta), \phi_2(\theta)) 
 \quad\mbox{in}\quad \Re
\theta \geq 0;
\hh  i=1, 2, 
 \ena
where $g^i$ is the generating function of $\hm_1$ if 
$\hm_0 = {\bf e}^{(i)}$:
\beas
g^i(s_1, s_2) = \sum_{r_1, r_2=0}^\infty p^i(r_1, r_2) s_1^{r_1}  s_2^{r_2}, 
\enas
where $ p^i(r_1, r_2) $ is the probability that an individual 
of type~$i$ has
$r_1$ children of type~1 and $r_2$ children of type~2.
Here, from the binomial structure, 
 \beas
g^1(s_1, s_2)&=& s_2 \left(\frac{\s}{\L} s_1 + 1 - \frac{\s}{\L}\right)^\L 
\enas
and 
\beas
g^2(s_1, s_2)&=& s_2 \left(\frac{\s}{\L} s_1 + 1 - \frac{\s}{\L} 
\right)^{2k\L}. 
 \enas 
The Laplace transforms $\phi_{k,\s}$ can be bounded as follows.

\begin{lemma} \label{dominate}
For $\th,\s > 0$, we have  
\beas
\phi_{k,\s; 1}(\theta) =: \phi_1(\theta) &\le& \frac{1}{1+\theta};\\
\phi_{k,\s; 2}(\theta) =:\phi_2(\theta) &\le& \frac{1}{1+\theta (\lambda - \s)}, 
\enas 
and hence 
\beas
\EE \left\{ e^{ - \theta \left(f\ui_1\right)^{-2} W_{k,\s} W'_{k,\s}}
  \giv \hm(0)=\hm'(0)={\bf e}\ui \right\}
   &\le& \theta^{-1} \log \left( 1 + \theta \right) . 
\enas 
\end{lemma}

\proof
We proceed by induction. Put
\beas
\phi_{i,n}(\theta) &=& \EE \left(e^{-\theta (f_1^{(1)})^{-1} W^{(1)}(n)} 
  \giv {\hat M}(0) = {\bf e}^{(i)} \right), \quad i=1,2. 
\enas
Then
\beas
\phi_{1,0} (\theta) &=& e^{-\theta} \le  \frac{1}{1+\theta};\\
\phi_{2,0}(\theta) &=& e^{-\theta(\lambda - \s)} 
    \le  \frac{1}{1+\theta (\lambda - \s)}.
\enas 
Assume that 
\beas
\phi_{1,n} (\theta) &\le&  \frac{1}{1+\theta};\\
\phi_{2,n}(\theta) & \le&  \frac{1}{1+\theta (\lambda - \s)}.
\enas 
By the Poincar\'{e} recursion, 
\beas
\phi_{i,n+1}(\theta)
&=& g^i\left( \phi_{1,n}\left(\frac{\theta}{\lambda}\right), 
    \phi_{2,n}\left( \frac{\theta}{\lambda}\right)\right) 
\enas 
for $i=1,2$. Hence, using the induction assumption, 
\beas
\phi_{1,n+1} (\theta) &\le& \frac\l{\l+\th(\l-\s)}
   \exp\Blb \s\Bl \frac\l{\l+\th} - 1 \Br\Brb\\
&\le& \frac\l{\l(1+\th)-\th\s}\,\frac{\l+\th}{\l+\th+\th\s}\\ 
&=& \frac{\l(\l+\th)}{\l(1+\th)(\l+\th) + \th^2(\l-1-\s)\s}\\   
&\le&  \frac{1}{1+\theta},
\enas 
and, also from~\Ref{ev},
\beas
\phi_{2,n+1} (\theta) &\le& \frac\l{\l+\th(\l-\s)}
   \exp\Blb 2k\s\Bl \frac\l{\l+\th} - 1 \Br\Brb\\
&\le& \frac\l{\l+\th+\th(\l-\s-1)}\,\frac{\l+\th}{\l+\th+2k\s\th}\\
&=& \frac\l{\l+\th+\th(\l-\s-1)}\,\frac{\l+\th}{\l+\th+(\l-1)(\l-\s)\th}\\
&=& \frac{\l(\l+\th)}
   {\l(\l+\th)(1+\th(\l-\s)) + \th^2(\l-1-\s)(\l-\s)(\l-1)}\\
&\le&  \frac{1}{1+\theta(\lambda - \s)}.
\enas 
Taking limits as $n\rightarrow \infty$ proves the first two assertions. 
The last assertion follows as in Lemma \ref{D-tail}.
\ep

\begin{theorem}\label{corollary6d} 
Let $\Delta_d$ denote a random variable on the integers with 
distribution given by
\begin{equation}\label{D-def}
\PP\left[\Delta_d > x \right]
= 
\EE\exp  \Blb -\messls
     \f_d^2 \l^x W_{k,\s} W'_{k,\s}\Brb ,
  \quad x \in \ZZ,
\en
and let $D^* = \D_d + 2n_d  $. Then
\beas
\dtv({\cal L}(D), {\cal L}(D^*)) = O\Bl \log(\L\s)(\L\s)^{-\g/(4-\g)} \Br,
\enas
uniformly in $\L$, $k$ and~$\s$ such that $k\s \le \r_0$, for any fixed
$0<\r_0 < \infty$, where $\g$ is given as in~(\ref{gamma-def}).  Hence
$\dtv({\cal L}(D), {\cal L}(D^*))\rightarrow 0$ if $\L\s \to \infty$
and $k\s$ remains bounded.
In particular, $\g=1/2$ if $k\s \le 1$, and the approximation error
is then of order $O(\log(\L\s)(\L\s)^{-1/7})$.
\end{theorem}

\proof
Fix~$G$, and consider~$x$ satisfying 
$x \le \left\lfloor \frac{G \log (\L \s) - 2 \log \phi_d}{\log \l} \right\rfloor$;
set $r(x) = \lfloor x/2 \rfloor$, $r'(x) = x - r(x) \le (x+1)/2$. 
Then it follows  from Theorem \ref{D4-a} and (\ref{kappabound}) that 
\beas
\lefteqn{\left\vert \PP[ D > 2 n_d + x]  - \PP[D^*> 2 n_d + x]
        \right\vert}\\
&\le&   (\h'_1(r(x),r'(x)) + \h'_2(r(x),r'(x)))(\L\s)^{-1/2}\\
&&\qquad  + \h'_3(r(x),r'(x))(\L\s)^{-1/4}
              + \h'_4(r(x),r'(x))(\L\s)^{-\g/2}\\
&=& O\left( \l^{1/2} \left( \frac{ k \s \log(\L\s)}{\log \l}\right)^2
         (\L\s)^{-(1-3G)/2} \right. \\
&&\qquad\mbox{}
  +  \left. \l^{1/2} \Bl\frac{ k \s \log(\L\s)}{\log \l} \right) 
         (\L\s)^{-(1-G)/2} \right. \\
&&\qquad\mbox{} + \left. \l^{3/4} (\l - \s)^2  (\L\s)^{-(1-3G)/4}\right. \\ 
&&\qquad\mbox{}
  + \left. \l^{1 - \g/2 }\left(\frac{\log (\L\s)}{\log \l} + 
      \left(\s^2 \vee (\l - 1)\right) \right)^{1/2}
           (\L\s)^{-(\g-G(2-\g))/2}\Br .  
\enas  
Also, from Lemma \ref{dominate}, recalling that $(f_1\ui)^{-2} = \l-\s$,
we have the upper tail estimate 
\beas
\lefteqn{\PP(\Delta_d > x) }\\
&\le&  \frac{(\l - \l_2)^2}{\l^2 (2\l - \s)} 
\log\left(1 +  \frac{\l^2}{(\l - \l_2)^2} (2\l - \s)(\L \s)^G \right) 
      (\L \s)^{-G} \\
&=&  O\left(\log(\L \s)(\L\s)^{-G}\Br.
\enas 
Comparing the exponents of $\L\s$, and remembering that $\g\le 1/2$,
the best choice of~$G$ is $G=\g/(4-\g)$, making $G = (\g-G(2-\g))/2$;
noting also that $\l = O(1+\s+\sqrt{k\s})$, the theorem is proved. 
\ep

Remembering that the choices $k\s=\r$ and $\L=Lk$ match
this model with that of Section~\ref{Sect2}, we see that $\L\s = L\r$,
and that thus Theorem~\ref{corollary6d} matches Theorem~\ref{corollary6}
closely for $\r\le1$, but that the total variation distance estimate
here becomes bigger as~$\r$ increases. Indeed, if $\r\to\infty$
and $\s = O(k)$,
then $\g(k,\s)\to0$, and no useful approximation is obtained. This
reflects the fact that, when $|\l_2|/\l$ is close to~$1$, the
martingale~$W\ui(n)$ only slowly comes to dominate the behaviour
of the two type branching process; for example, from~\Ref{one-star},
$$
\hm^+(n) =\frac{1}{\lambda - \lambda_2}\left( 
\frac{\lambda^{n+1}}{f_1^{(1)}} W^{(1)}(n) -
\frac{\lambda_2^{n+1}}{f_1^{(2)}} W^{(2)}(n)\right)  
$$
then retains a sizeable contribution from $W\ut(n)$ until~$n$ becomes
extremely large.  This is in turn a consequence of taking the
shortcuts to have length~$1$, rather than~$0$; as a result,
the big multiplication,
by a factor of~$2\r$, occurs only at the {\it second\/} time step,
inducing substantial fluctuations of period~$2$ in the
branching process, which die away
only slowly when~$\r$ is large.
However, if $\r\to\infty$ and $k = O(\s^{1-\e})$ for any $\e > 0$,
then $\liminf\g(k,\s) > 0$, and it becomes possible for $\law(D)$ 
and~$\law(D^*)$ to be asymptotically close in total variation.
This can be deduced from the proof of the theorem by taking
$k\sim L^\a$ and $\s\sim L^{\a+\b}$, 
for choices of $\a$ and~$\b$ which ensure  
that $\s^2$ dominates~$\r$. Under such circumstances, the
effect of two successive multiplications by~$\s$ in the branching
process dominates that of a single multiplication by~$2\r$ at the
second step, and approximately geometric growth at rate $\l\sim\s$
results.  However, as in all situations in which~$\r$ 
is a positive power of~$\L$, interpoint distances are asymptotically
bounded, and take one or at most two values with very high
probability; an analogue of Corollary~\ref{rholarge} could for
instance also be proved.

If $\rho = k\s$ is small, we can again compare the distribution
of~$W_{k,\s}$ with the NE$(1)$ distribution of the limiting 
variable~$W$ in the Yule process (see \cite{BR}), using the fact
that its Laplace transforms satisfy the Poincar\'e 
equation~\Ref{Poincare}.  Define the operator~$\Xi$ by 
\beas
(\Xi \phi)_1(\theta) 
&:=& g^1\left(\phi_1\left(\frac{\theta}{\lambda}\right), 
   \phi_2\left(\frac{\theta}{\lambda}\right)  \right) \\
&=& \phi_2\left(\frac{\theta}{\lambda}\right) 
\left(\frac{\s}{\L}  \phi_1\left(\frac{\theta}{\lambda}\right)  
      + 1 - \frac{\s}{\L}  \right)^\L ;\\
(\Xi \phi)_2(\theta) 
&:=& g^2\left(\phi_1\left(\frac{\theta}{\lambda}\right) ,
\phi_2\left(\frac{\theta}{\lambda}\right)  \right)\\
&=& \phi_2\left(\frac{\theta}{\lambda}\right)
\left(\frac{\s}{\L}  \phi_1\left(\frac{\theta}{\lambda}\right) 
 + 1 - \frac{\s}{\L}  \right)^{2k\L}. 
 \enas 
Let 
$$ 
{\cal G} := \Blb\g=(\g_1, \g_2): [0, \infty)^2 \rightarrow \RR: 
  \| \g \|_{\cal G}  :=  \sup_{\theta > 0} 
 \max \left\{\frac{|\g_1(\theta)|, |\g_2 (\theta)|}{\theta^2} \right\} 
  < \infty\Brb,
$$
and 
\beas 
{\cal H} &:= &\Blb \psi=(\psi_1, \psi_2): [0, \infty)^2 \rightarrow \RR:
\frac{\psi_1(\theta) -( 1 - \theta )}{\theta^2} \mbox{ is bounded}, 
            \right.\\
&&\qquad\left. \frac{\psi_2(\theta) -( 1 - \theta(\lambda - \s))}{\theta^2}  
 \mbox{ is bounded} \Brb.
\enas  
Then ${\cal H}$ contains $\f_{k,\s} = (\f_1,\f_2)$ as defined
in~(\ref{phi}), since  
$$
\EE \{(f_1^{(1)})^{-1} W_{k,\s} \giv \hm(0) = {\bf e}\ui\} = 1;
\quad  \EE \{(f_1^{(1)})^{-1} W_{k,\s} \giv \hm(0) = {\bf e}\ut\} 
   = \l-\s,
$$
and taking limits in (\ref{inc1}) shows that $\var W_{k,\s}$ exists. 
We next show that~$\Xi$ is a contraction on~${\cal H}$. 

\begin{lemma} \label{contd}
The operator $\Xi$ is a  contraction  on ${\cal H}$, and, for all 
$\psi, \chi \in {\cal H}$,  
\beas
\| \Xi  \psi -  \Xi  \chi \|_{\cal G} &\leq& \Bl\frac{2k\s+
1}{\lambda^2}\Br \| \psi - \chi \|_{\cal G}. 
 \enas
\end{lemma}

\remark  Note that
\beas
\frac{2k\s+1}{\lambda^2} 
&=& \frac{\lambda^2 - (\lambda-1) (\s + 1)}{\lambda^2} < 1.  
\enas 

\proof
For all $\psi, \chi \in
{\cal H}$ and $\theta > 0$,  observe that $\psi - \chi \in {\cal G}$. 
We then compute
\beas
\left| (\Xi  \psi)_1(\theta) - (\Xi  \chi)_1(\theta)\right|  &\leq &
\left| \psi_2 \left(\frac{\theta}{\lambda}\right) - \chi_2   
\left(\frac{\theta}{\lambda}\right) \right| 
\\
&&+ \s \left| \psi_1 \left(\frac{\theta}{\lambda}\right) - 
\chi_1   \left(\frac{\theta}{\lambda}\right) \right|, 
\enas 
so that
\bea \label{*}
\frac{\left| (\Xi  \psi)_1(\theta) - (\Xi  \chi)_1(\theta)\right| 
}{\theta^2} &\leq&
\frac{1}{\lambda^2} \frac{ \left| \psi_2
\left(\frac{\theta}{\lambda}\right) - \chi_2   
\left(\frac{\theta}{\lambda}\right) \right|}{\left(
\frac{\theta}{\lambda}\right)^2}  \nonumber  \\ 
&&+ \frac{\s}{\lambda^2} \frac{
\left| \psi_1 \left(\frac{\theta}{\lambda}\right) -  \chi_1  
\left(\frac{\theta}{\lambda}\right) \right|}{\left(
\frac{\theta}{\lambda}\right)^2}\nonumber \\
&\leq& \frac{\s + 1}{\lambda^2}\, \| \psi - \chi \|_{\cal G}.   
\ena 
Similarly,
\beas
\left| (\Xi  \psi)_2(\theta) - (\Xi  \chi)_2(\theta)\right|  &\leq &
\left| \psi_2 \left(\frac{\theta}{\lambda}\right) - \chi_2 
 \left(\frac{\theta}{\lambda}\right) \right| 
\\
&&+ 2k\s \left| \psi_1 \left(\frac{\theta}{\lambda}\right) - 
\chi_1   \left(\frac{\theta}{\lambda}\right)  \right|,
\enas
and
\beas  
\frac{\left| (\Xi  \psi)_2(\theta) - (\Xi  \chi)_2(\theta)\right| 
}{\theta^2} &\leq& 
\Bl\frac{2k \s + 1}{\lambda^2}\Br\, \| \psi - \chi \|_{\cal G}.  
\enas 
Taking the maximum of the bounds finishes the proof. 
\ep 

Thus, for any starting function $\psi = (\psi_1, \psi_2) \in {\cal H}$ 
and for $\phi_{k,\s} = (\phi_1 , \phi_2) $ given in (\ref{phi}), 
we have 
\beas
\| \phi_{k,\s}  - \psi \| _{\cal G} &\leq & \| \Xi \phi_{k,\s} -
  \Xi \psi  \| _{\cal G} + \| \Xi \psi  - \psi \| _{\cal G} \\
&\leq& 
\frac{2k\s + 1}{\lambda^2}  \,\| \phi_{k,\s}  - \psi \| _{\cal G} 
   + \| \Xi \psi  - \psi \| _{\cal G},
\enas 
so that
\bea \label{contbound}
\| \phi_{k,\s}  - \psi \| _{\cal G} &\leq &
\frac{\lambda^2}{\lambda^2-(2k\s + 1)} \, \| \Xi \psi  - \psi \| _{\cal
G} . 
\ena 
Hence a function $\psi $ such that $\| \Xi \psi  - \psi \|_{\cal G}$ 
is small provides a good approximation to~$\phi_{k,\s}$. 

As a candidate $\psi$, we try
\bea 
 \psi_{(1)}(\theta) &=& \frac{1}{1+ \theta}, \non\\
\psi_{(2)}(\theta) &=& \frac{1}{1+ \theta(\lambda - \s)} \label{psi}; 
\ena
Lemma \ref{dominate} shows that this pair 
dominates~$\phi_{k,\s}$. 

\begin{lemma} \label{psilemma}
For $\psi$ given in (\ref{psi}), we have
$$
\|\Xi\ps - \ps\|_{{\cal G}} \le 
   \frac{2k\s(\lambda^2 - \lambda \s - 1 +  k \s)}{\lambda^2}.
$$
\end{lemma}

\ignore{    
\beas
\frac{\left| (\Xi  \psi)_1(\theta) - \psi_1(\theta)\right| }{\theta^2} 
 &\leq& \frac{2\s(\lambda - 1 + \lambda \s)}{2\lambda^2}, 
\enas 
and 
\beas
\frac{\left| (\Xi  \psi)_2(\theta) - \psi_2(\theta)\right| }{\theta^2} 
 &\leq& \frac{2k\s(\lambda^2 - \lambda \s - 1 +  k \s)}{\lambda^2}.
\enas 
\end{lemma}
}

\proof
For $\theta > 0$, we have
\beas
\lefteqn{(\Xi \psi)_1  (\theta) - \psi_1 (\theta)}\\
&=& \frac{\lambda}{\lambda + \th (\lambda - \s)} 
  \left( 1 - \frac{\s \th}{\L(\lambda + \th)}\right)^\L - \frac{1}{1+ \th}\\
&=& \frac{1}{1+ \th}\left\{ 
  \frac{\lambda(1+ \th) }{\lambda + \th (\lambda - \s)} 
    \left( 1 - \frac{\s \th}{(\lambda + \th)}\right) -1 \right\} +R_1, 
\enas
where
\beas
R_1 &=& \frac{\lambda}{\lambda + \th (\lambda - \s)}
  \left[\left( 1 - \frac{\s \th}{\L(\lambda + \th)}\right)^\L 
     -1 +\frac{\s \th}{\lambda + \th} \right] .
\enas 
Moreover,
\beas
\lefteqn{\frac{1}{1+ \th}\left\{  
 \frac{\lambda(1+ \th) }{\lambda + \th (\lambda - \s)} 
   \left( 1 - \frac{\s \th}{\lambda + \th}\right) -1 \right\}}\\
&& = \frac{\theta^2 \s(1 - \lambda)}
    {(1+ \th)(\lambda + (\lambda - \s)\th)(\lambda + \th)}. 
\enas 
From Taylor's expansion, it follows that 
\beas
\vert R_1 \vert 
&\le& \frac{\l\L(\L-1) \s^2 \th^2}
   {2(\lambda + (\lambda - \s)\th)\,\L^2 (\lambda + \th)^2}\\
&\le& \frac{\s^2 \th^2}{2 \lambda^2}. 
\enas 
Hence
\bea \label{***}
\frac{\left|(\Xi  \psi)_1(\theta) - \psi_1(\theta)\right| }{\theta^2} 
  &\leq& \frac{\s(2(\lambda - 1) + \s)}{2\lambda^2}.
\ena 
Similarly, 
\beas
\lefteqn{(\Xi \psi)_2  (\theta) - \psi_2 (\theta)}\\
&=& \frac{\lambda}{\lambda + \th (\lambda - \s)} 
 \left( 1 - \frac{\s \th}{\L(\lambda + \th)}\right)^{2k\L} 
     - \frac{1}{1+ \th(\lambda -\s)}\\
&=& \frac{1}{1+ \th(\lambda -\s) }\left\{  
 \frac{\lambda(1+ \th(\lambda -\s) ) }{\lambda + \th (\lambda - \s)} 
  \left( 1 - \frac{2k \s \th}{\lambda + \th}\right) -1 \right\} +R_2, 
\enas
where
\beas
R_2 &=&  \frac{\lambda}{\lambda + \th (\lambda - \s)}
  \left[\left(1 - \frac{\s \th}{\L(\lambda + \th)}\right)^{2k\L}
     -1 +\frac{2k\s \th}{\lambda + \th} \right] .
\enas 
Using (\ref{ev}), we obtain  
\beas
\lefteqn{\frac{1}{1+ \th(\lambda - \s)}\left\{  
    \frac{\lambda(1+ \th(\lambda -\s))}{\lambda + \th (\lambda - \s)} 
      \left( 1 - \frac{2k \s \th}{\lambda + \th}\right)-1 \right\}}\\
&=& \frac{\theta^2 (\lambda -1)(\lambda -\s)(1 + \lambda \s - \lambda^2)}
 {(1+ \th(\lambda -\s))(\lambda + (\lambda - \s)\th)(\lambda + \th)}\\
&=& \frac{2 k \s \theta^2 (1 + \lambda \s - \lambda^2)}
 {(1+ \th(\lambda -\s))(\lambda + (\lambda - \s)\th)(\lambda + \th)}. 
\enas 
From Taylor's expansion, it now follows that 
\beas
\vert R_2 \vert &\le& 
\frac{2k\L(2k\L-1)\l \s^2 \th^2}
    {2(\lambda + (\lambda - \s)\th)\,\L^2 (\lambda + \th)^2}\\
&\le& \frac{2 k^2 \s^2 \th^2}{ \lambda^2}. 
\enas 
Hence
\beas
\frac{\left| (\Xi  \psi)_2(\theta) - \psi_2(\theta)\right| }{\theta^2} 
 &\leq& \frac{2k\s(\lambda^2 - \lambda \s - 1 + k \s)}{\lambda^2},
\enas 
completing the proof, since
$$
2(\lambda^2 - \lambda \s - 1 + k \s) = 2(\l-1) + 2(3k-1)\s
  > 2(\l-1) + \s.
$$
\ep

This enables us to prove the exponential approximation to~$\law(W_{k,\s})$
when~$k\s$ is small. 

\begin{theorem} \label{drhosmall} 
As $k\s \to 0$, $\law(W_{k,\s}) \to 
{\rm NE}(1)$.
\end{theorem} 

\proof
Let $\phi_{k,\s}$ be as in~\Ref{phi}, and~$\ps$ as in~\Ref{psi}.
Then $(\phi_{k,\s})_1$ is the Laplace transform of
$$
\law(\fiimi W_{k,\s}) := \law(\fiimi W_{k,\s} \giv \hm(0) = {\bf e}\ui),
$$ 
and~$\psi_1$ that
of NE$(1)$, and $\fiimi = (\l-\s)^{1/2} \to 1$ as
$k\s \to 0$.  Hence it is enough to show that
$$
\lim_{k\s\to0} \|\phi_{k,\s} - \ps\|_{{\cal G}} = 0.
$$
However, using Lemma~\ref{psilemma} and~(\ref{contbound}), we obtain 
\bea  \label{**}
&&\|\phi_{k,\s} - \ps\|_{{\cal G}} \nonumber \\
&&\qquad\le \Bl\frac{\l^2}{\l^2-1-2k\s}\Br
                  \|\Xi\ps - \ps\|_{{\cal G}} \nonumber \\
&&\qquad\leq 2k \s \,\frac{\lambda^2 - \lambda \s - 1 +  k \s}
    {\lambda^2 - 1 - 2k\s}  \nonumber \\
&&\qquad\leq 2k\s\,\frac{k\s(5+2\s)}{2k\s(\s+1)}\,
   (1+2k\s(\s+1)) \nonumber \\
&&\qquad\leq k\s(1+2k\s)(5+2\s)    \to 0, 
\ena 
since $k\s\to0$.
This proves the theorem. \ep

Again we can use this result to derive an approximation to the
distribution of the distance for $D$, based on a corresponding 
distribution derived from the NE(1) distribution. The starting point
is the following result.

\begin{theorem}\label{3.14analog}
Let $W,W'$ be independent NE(1) variables. Then, for all $\theta > 0$, 
\beas
\lefteqn{ 
\left \vert \EE\exp  \Blb -\theta(\l-\s) W_{k,\s} W'_{k,\s}\Brb 
- \EE e^{-\theta W W'}\right\vert}\\ 
&\le& 5\theta^2 k\s \left\{ (3\s/2) + ( 1 + 2k\s )(5+2\s) \right\} . 
\enas 
\end{theorem}

\proof
As in the proof of Theorem~\ref{comp}, with  
$\phi_{k,\s}$ as in~\Ref{phi} and~$\ps$ as in~\Ref{psi}, and because,
from~\Ref{f11-def}, $\l-\s = (\fii)^{-2}$, we have
\beas
\lefteqn{\EE\exp  \Blb -\theta(\l-\s) W_{k,\s} W'_{k,\s}\Brb 
- \EE e^{-\theta W W'}}
\\
&=& \EE \Blb(\phi_{k,\s})_1( \theta \fiimi W'_{k,\s}) \Brb
- \EE \psi_1(\theta W )\\
&=& \EE \Blb(\Xi \phi_{k,\s})_1( \theta\fiimi  W'_{k,\s})\Brb 
- \EE\Blb (\Xi \psi)_1(\theta\fiimi  W'_{k,\s} )\Brb\\
&&+ \EE \Blb(\Xi \psi)_1(\theta  \fiimi W'_{k,\s} )\Brb 
      - \EE \Blb\psi_1(\theta \fiimi W'_{k,\s} )\Brb\\
&&+ \EE \Blb\psi_1(\theta \fiimi W'_{k,\s} )\Brb -  \EE \psi_1(\theta W ). 
\enas 
Now~\Ref{*}  in the proof of Lemma~\ref{contd} gives 
\beas
\lefteqn{
{\left\vert  \EE \Blb(\Xi \phi)_1( \theta \fiimi W'_{k,\s})\Brb 
- \EE\Blb (\Xi \psi_{k,\s})_1(\theta \fiimi W'_{k,\s} )\Brb \right\vert }}\\
&\le & \theta^2 (\fii)^{-2}\frac{\s+1}{\l^2}\, 
       \| \phi_{k,\s} - \psi\|_{\cal G} \ex\{(W'_{k,\s})^2\}\\
&\le& \theta^2 (1+\k^2)\frac{\s+1}{\l^2}\, 
       \| \phi_{k,\s} - \psi\|_{\cal G} \\ 
&\le& 3\theta^2 \frac{\s+1}{\l^2}\, 
       \| \phi_{k,\s} - \psi\|_{\cal G} ,
\enas
from \Ref{wrho}, \Ref{inc1} and~\Ref{kappabound},
and~\Ref{**} then implies that the above 
expression can be bounded by
\beas
3\theta^2\, \frac{\s+1}{\l^2}\, k \s (1 + 2 k \s)(5 + 2 \s).
\enas 
Similarly, from ~\Ref{***} in the proof of Lemma~\ref{psilemma}, 
\beas
{\left\vert \EE \Blb(\Xi \psi)_1(\theta \fiimi W'_{k,\s} ) \Brb
      - \EE \Blb\psi_1(\theta \fiimi W'_{k,\s} ) \Brb \right\vert}
&\le& 3\theta^2 \,\frac{\s \{2 (\l -1) + \s \} }{2 \l^2}.
\enas 
Then, with $W \sim \NE(1)$ independent of $W'_{k,\s}$, we have
$$  
\EE \Blb \psi_1(\theta \fiimi W'_{k,\s} ) \Brb 
   = \EE \Blb e^{-\theta \fiimi W W'_{k,\s}}\Brb  
     = \EE\Blb (\phi_{k,\s})_1(\theta W)\Brb , 
$$ 
and hence,
from~\Ref{**} in the proof of Theorem~\ref{drhosmall}, it follows that
\beas
{\left\vert \EE \Blb\psi_1(\theta\fiimi  W'_{k,\s} )\Brb 
    -  \EE \psi_1(\theta W ) \right\vert}
&=& \left\vert \EE \Blb(\phi_{k,\s})_1(\theta W)\Brb  
                 - \EE \psi_1(\theta W) \right\vert \\
&\le& 2\theta^2 k \s (1 + 2 k \s)(5+2 \s). 
\enas 
Since $\l-1 < 2k\s $ and $\l^2 > \s+1$, the assertion follows from the 
triangle inequality. 
\ep

Recalling that 
$$
\EE e^{- \theta WW'} = \int_0^\infty {e^{-y} \over 1+\th y}\,dy, 
$$
we can now derive the analogue of Theorem \ref{corollary14}.

\begin{theorem} \label{corollary14d}
As in Theorem \ref{corollary6d}, let 
$\Delta_d$ denote a random variable on the integers with 
distribution given by
\beas
\PP\left[\Delta_d > x \right]
= 
\EE\exp  \Blb -\messls
     \f_d^2 \l^x W_{k,\s} W'_{k,\s}\Brb ,
  \quad x \in \ZZ. 
\enas
Let $T$ denote a random variable on~$\RR$ with
distribution given by  
\beas
\PP\left[T > z\right] = \int_0^\infty \frac{e^{-y}}{1 + 2ye^{2z}}\, dy.
\enas 
Then 
\beas 
\sup_{z \in \RR} \left\vert \PP\left[\lmit \Delta_d > z\right] 
                 - \PP[T  > z]\right\vert 
 &=& 
   O\Blb (k\s)^{1/3} \left( 1 + \log \left (1/k\s\right)\right) \Brb,
\enas 
uniformly in $k\s \le 2$.
\end{theorem}

\proof
 We use an argument similar to the proof of Theorem~\ref{corollary14}. 
Putting 
\beas
\ctilde(\l) = \frac{\log\l}{\l-1} 
  = \frac{\log( 1 + 2 \,\frac{\l-1}{2} )}{ 2 \,\frac{\l-1}{2} },
\enas 
we have, as before, 
$$
1 \ge \ctilde (\l) \ge 1 - \frac{\l-1}{2};
$$
we also write 
\beas
\b(\l) &:=& \left(\frac{\l}{\l-\l_2}\right)^2 
           \left( \l - \frac{\s}{2}\right) \phi_d^2.
\enas

We use the following bounds. First, from the characteristic equation~\Ref{ev}, 
we have
\beas
0\ \le\ \l-1 &= &\frac{2k\s}{\l-\s}\ \le\ 2 k \s.
\enas
Then, since $\sqrt{a+b} \le \sqrt{a} + \sqrt{b}$ for $a,b \ge 0$, 
it follows that
\beas
1 + \s \ \le\ \l &\le& 1 + \s + \sqrt{\s (2k-1)}.
\enas
Thus, and because $\l^{-1} \le \phi_d \le 1$ and $\l_2 < 0$, we have
\beas
\b(\l) &\le& 1 + (\s/2) + \sqrt{\s(2k-1)} 
\enas
and
\beas
\b(\l) &\ge& \left( \frac{1}{\l - \l_2}\right)^2\\
&=& 1 - \frac{2 \s(4k-1) + \s^2}{(\s+1)^2 + 4 \s(2k-1)} .
\enas
This in turn gives
\beas
|\b(\l) -1| &\le& \max\left\{ (\s/2) + \sqrt{\s(2k-1)},
\frac{{2\s(4k-1)+\s^2}}{(\s+1)^2 + 4 \s(2k-1)} \right\}\\
 &=:& \Gamma(\s, k) \\
 &=& O\left( \max\{ \s, \sqrt{k\s} \}\right).
\enas

For the main part of the distribution, 
we write
\bea 
\lefteqn{\PP\left[\lmit \Delta_d > z\right] - \PP[T  > z]}\nonumber \\
&=& \PP\left[\lmit \Delta_d > z\right] 
 - \int_0^\infty e^{-y} \left( 1 + 2 y \b(\l)e^{2z\ctilde(\l)}\right)^{-1}dy 
         \label{approx1}\\
&&\mbox{}+ \int_0^\infty e^{-y} 
       \left( 1 + 2 y \b(\l)e^{2z\ctilde(\l)}\right)^{-1}dy - 
       \PP[T > z\ctilde(\l)]\label{approx2}\\
&&\mbox{}+ \PP[T > z\ctilde(\l)] -   \PP[T > z] \label{approx3}.
\ena
Now, for ~\Ref{approx1}, Theorem \ref{3.14analog} yields
\beas
\lefteqn{\left\vert \PP\left[\lmit \Delta_d > z \right] -
 \int_0^\infty e^{-y} \left( 1 + 2 y \b(\l)e^{2z\ctilde(\l)}\right)^{-1} \,dy
  \right\vert }\\
&=&
\left\vert \EE\exp  \Blb - 2 \b(\l) e^{2z\ctilde(\l)} W_{k,\s} W'_{k,\s}\Brb -
 \int_0^\infty e^{-y} \left( 1 + 2 y \b(\l)e^{2z\ctilde(\l)}\right)^{-1} \,dy 
  \right\vert \\
&\le&
 4\b(\l)^2 e^{4z\ctilde(\l)} 
   5k\s \left\{(3\s/2) + ( 1 + 2k\s )(5+2\s)\right\}\\
&\le& (2\l-\s)^2 e^{4z} 
   5k\s \left\{(3\s/2) + ( 1 + 2k\s )(5+2\s)\right\}.
\enas
With ~\Ref{star-d}, we have, for~\Ref{approx2}, that
\beas
\lefteqn{\left\vert\int_0^\infty e^{-y} 
  \left( 1 + 2 y \b(\l)e^{2z\ctilde(\l)}\right)^{-1} \,dy
     - \PP[T > z\ctilde(\l)]\right|}\\
&\le& 2 e^{2z\ctilde(\l)} {|\b(\l)-1| 
  \over \max\{1, 2\b(\l) e^{2z\ctilde(\l)}, 2 e^{2z\ctilde(\l)}\}} \\
&\le& 2 { e^{2z\ctilde(\l)} \over \max\{1, 2 e^{2z\ctilde(\l)}\}}\,|\b(\l)-1|\\
&\le&  \Gamma(\s,k).
\enas
Similarly, for~\Ref{approx3}, because $1 - \ctilde(\l) \le \frac{\l-1}{2} \le
 k \s $ and from Taylor's expansion,
it follows that
\beas
\lefteqn{\left\vert \PP[T > z \ctilde (\l)] - \PP[T > z] \right\vert}\\
&=& \left\vert \int_0^\infty e^{-y} 
  \left( 1 + 2 y e^{2z\ctilde(\l)}\right)^{-1} \,dy 
    - \int_0^\infty e^{-y} \left( 1 + 2 y e^{2z}\right)^{-1} \,dy \right\vert\\
&\le&  2 e^{2z}{|e^{-2z(1-\ctilde(\l))}-1| 
     \over \max\{1, 2  e^{2z\ctilde(\l)}, 2 e^{2z}\}}.
\enas
If $z>0$, this gives
$$
|\PP[T > z \ctilde (\l)] - \PP[T > z]| \le 2z(1-\ctilde (\l)) \le 2k\s z;
$$
if $z \le 0$, we have
\beas
|\PP[T > z \ctilde (\l)] - \PP[T > z]| &\le& 2|z|(1-\ctilde (\l))
     e^{-2z\ctilde (\l)} \ \le\ 2k\s|z| e^{2z(1-k\s)}.
\enas
Hence we conclude that, uniformly in $k\s \le 1/2$,
\bea
\lefteqn{\PP\left[\lmit \Delta_d > z\right] - \PP[T  > z]}\nonumber \\
&\le& 5k\s e^{4z} (2\l-\s)^2 \left\{(3\s/2) + ( 1 + 2k\s )(5+2\s)\right\}\non\\
&&\quad\mbox{} + \Gamma(\s,k) + 2k\s|z| \min\{1,e^{2z(1-k\s)}\}\non\\
&\le& C_1\Blb k\s(e^{4z}+1) + \sqrt{k\s} \Brb, \label{E1}
\ena
for some constant~$C_1$.

For the large values of~$z$, where the bound given in~\Ref{E1} becomes
useless, we can estimate the upper tails of the random variables
separately. First, for $x\in\ZZ$, we have
\beas
\PP\left[\Delta_d > x \right] =
   \EE\exp  \Blb - 2 \b(\l) \l^x (\fii)^{-2}W_{k,\s} W'_{k,\s}\Brb ,
\enas
so that, by Lemma~\ref{dominate}, it follows that 
\beas
\lefteqn{\PP\left[\lmit \Delta_d > z \right]}\\
&=& \EE\exp  \Blb 
    - 2 \b(\l) e^{2z\ctilde(\l)}(\fii)^{-2} W_{k,\s} W'_{k,\s}\Brb \\
&\le& \left(\frac{\l-\l_2}{\l} \right)^2 (2\l-\s)^{-1} \phi_d^{-2} 
      e^{-2z \ctilde(\l)} 
   \log \left(1 + \left( \frac{\l}{\l-\l_2} \right)^2 (2\l-\s) \phi_d^2 
       e^{2z\ctilde(\l)} \right)
\\
&\le& 4\l e^{-2z \ctilde(\l)} 
  \log \left(1 + (2\l-\s) e^{2z\ctilde(\l)} \right) \\
&\le& 4\l e^{-2z\left(1 - k\s\right) } \log \left(1 + 
     \left( 2 + \s + 2 \sqrt{\s(2k-1)} \right) e^{2z} \right),
         \qquad z \in \lmit \ZZ.
\enas
For the upper tail of $T$, as in~\Ref{Tzcrho} and with $z>0$, 
we have 
 \beas
\PP[T > z\ctilde (\l)] &\leq&   e^{-2z\ctilde (\l)}(1+z\ctilde (\l)) 
   \ \le\ (1+z )e^{-2z\left( 1 - k\s \right) }.
\enas  
Combining these two tail estimates, we find that, for $z > 0$,
\bea
 \left\vert \PP [T > z \ctilde (\l)] - \PP[T > z] \right\vert 
&\le& C_2 (1+z)e^{-2z(1- \s k)},\label{E2}
\ena
uniformly in $k\s \le 1/2$, for some constant~$C_2$. 
Applying the bound~\Ref{E1} when $z\le (6-2k\s)^{-1}\log(1/k\s)$ 
and~\Ref{E2} for all larger~$z$, and remembering that~$T$ has
bounded density, so that the discrete nature of~$\D_d$ requires
only a small enough correction, a bound of the required order
follows.
\ep

\bigskip
{\bf Acknowledgement.} We would like to thank Svante Janson for very helpful
discussions concerning the comparison with \cite{NewmanWatts2}.

\vfil\eject
\end{document}